\documentclass[12pt, draftclsnofoot, onecolumn]{IEEEtran}
\usepackage{setspace}
\usepackage[utf8]{inputenc}
\usepackage{amsmath}
\usepackage{amsfonts}
\usepackage{amssymb}
\usepackage{algpseudocode}
\usepackage{algorithm}
\usepackage[pdftex]{graphicx}
\usepackage{epstopdf,epsfig}
\usepackage{enumerate}
\usepackage{pgfplots}
\usepackage{subfigure}
\usepackage{mathtools}
\usepackage{amsthm}
\usepackage{wrapfig}
\usepackage{capt-of}
\usepackage[english]{babel}
\usepackage{url}
\usepackage{color}
\usepackage{xcolor}
\usepackage{changes}

\ifCLASSOPTIONcompsoc
  \usepackage[nocompress]{cite}
\else
  \usepackage{cite}
\fi

\theoremstyle{remark}

\theoremstyle{theorem}

\newtheorem{theorem}{Theorem}[section]
\newtheorem{lemma}[theorem]{Lemma}
\newtheorem{corollary}[theorem]{Corollary}
\newtheorem{proposition}[theorem]{Proposition}

\theoremstyle{definition}
\newtheorem{definition}{Definition}[section]

\newcommand{\karmoose}{\color{black}}

\newenvironment{Pf}{\textbf{Proof:}}{\hfill$\blacksquare$}

\newcommand{\Set}{\mathcal}
\title{Privacy in Index Coding: $k$-Limited-Access Schemes}
\author{Mohammed~Karmoose,
	Linqi~Song,
        Martina~Cardone,
        and~Christina~Fragouli
\thanks{
M. Karmoose and C. Fragouli are with the Electrical and Computer Engineering Department, University of California, Los Angeles (UCLA), CA 90095 USA (e-mail: mkarmoose@ucla.edu, christina.fragouli@ucla.edu).
L. Song is with the Department of Computer Science at the City University of Hong Kong (email: linqi.song@cityu.edu.hk).
M. Cardone is with the Electrical and Computer Engineering Department of the University of Minnesota, MN 55404 USA (e-mail: cardo089@umn.edu). Most of the work in this paper occurred when L. Song and M. Cardone were at UCLA.
The work of the authors was partially funded by NSF under Awards 1527550, 1514531, 1423271, 1314937 and 1740047.\newline The results in this paper were presented in part at the 2017 IEEE Information Theory Workshop and the 2018 IEEE International Symposium of Information Theory.}}
\IEEEoverridecommandlockouts

\begin{document}

\maketitle
\begin{abstract}
~
In the traditional index coding problem, a server employs coding to send messages to $n$ clients within the same broadcast domain. Each client already has some messages as side information and requests a particular unknown message from the server.
All clients  learn the coding matrix so that they can decode and retrieve their requested data. 
Our starting observation is that, learning the coding matrix can pose privacy concerns: it may enable a client to infer information about the requests and side information of other clients.
In this paper, we 
mitigate this privacy concern by allowing each client to have limited access to the coding matrix. In particular, we design coding matrices so that each client needs only to learn some of (and not all) the rows to decode her requested message. 
By means of two different privacy metrics, we first show that this approach indeed increases the level of privacy.
Based on this, we propose the use of $k$-limited-access schemes: given an index coding scheme that employs $T$ transmissions, we create a $k$-limited-access scheme with $T_k\geq T$ transmissions, and with the property that each client needs at most $k$ transmissions to decode her message.
We derive upper and lower bounds on $T_k$ for all values of $k$, and develop deterministic designs for these schemes, which are universal, {\it i.e.}, independent of the coding matrix. 
We show that our schemes are order-optimal when either $k$ or $n$ is large.
Moreover, we propose heuristics that complement the universal schemes for the case when both $n$ and $k$ are small.

\end{abstract}

\section{Introduction}

It is well recognized that broadcasting can offer significant bandwidth savings compared to point-to-point communication \cite{el2011network,fragouli2006network}, and could be leveraged in several wireless network applications. 
Use cases include Wi-Fi (cellular) networks where an access point (a base station) is connected to a set of Wi-Fi (cellular) devices through a wireless broadcast channel, and where devices request messages, such as YouTube videos. 
Another use case has recently emerged in the context of distributed computing~\cite{li2018fundamental,ezzeldin2017communication}, where worker nodes exchange data among themselves to complete computational tasks.

A canonical setup which captures the essence of broadcast channels is the index coding  framework~\cite{bar2011index}. In an index coding instance, a server is connected to a set of clients through a noiseless broadcast channel. The server has a database that contains a set of messages. Each client: 1) possesses a subset of the messages that she already knows, which is referred to as the \textit{side information set}, and 2) requests a message from the database which is not in her side information set. The server has full knowledge of the requests and side information sets of all clients. A \textit{linear index code} (or {\it index code} in short)\footnote{In this work, we solely focus on linear index codes.} is a linear coding scheme that comprises a set of coded broadcast transmissions which allow each client to decode her requested message using her side information set. The goal is to find an index code which uses the smallest possible number of broadcast transmissions. The key ingredient in designing efficient ({\it i.e.}, with a small number of transmissions) index codes is the use of coding across messages. 

The starting observation of this work is  that, using coding over broadcast channels can cause privacy risks.
In particular, a curious client may infer information about the requests and side information sets of other clients, which can be deemed sensitive by their owners.
For example, consider a set of clients that use a server to download YouTube videos. Although YouTube videos are publicly available, a client requesting a video about a medical condition may not wish for others to learn  her request, or learn what are other videos that she has already downloaded.


\begin{figure}
 \centering
  \includegraphics{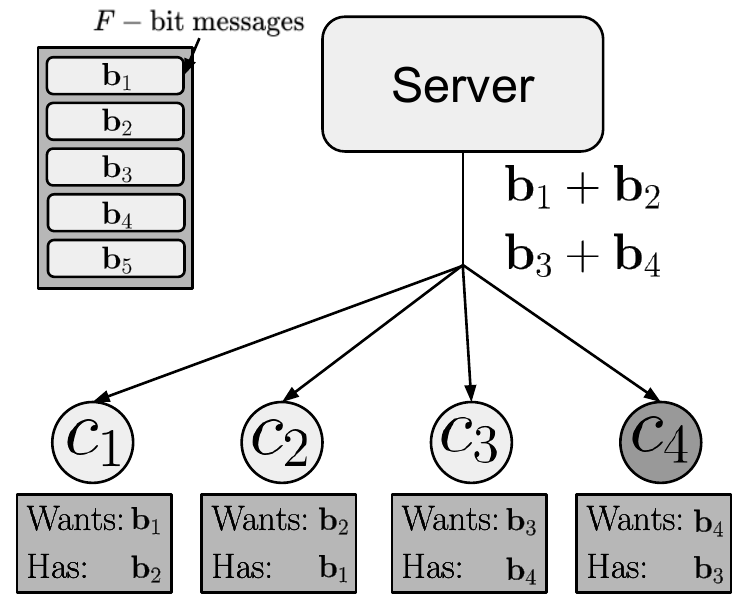}
 \caption{An index coding example with $5$ messages and $4$ clients. Each client wants one message and has another as shown above. The optimal index code consists of sending the two transmissions ${\bf b}_1 + {\bf b}_2$ and ${\bf b}_3 + {\bf b}_4$.}
 \label{fig::IndexCodingExample}
\end{figure}

To illustrate why coding can create privacy leakage,  consider the index coding instance shown in Figure~\ref{fig::IndexCodingExample}. A server possesses a set of $5$ messages, which we refer to as ${\bf b}_1$ to ${\bf b}_5$. 
The server is connected to a set of $4$ clients: client $1$ wants message ${\bf b}_1$ and has as side information message ${\bf b}_2$; client $2$ wants ${\bf b}_2$ and has ${\bf b}_1$; client $3$ wants ${\bf b}_3$ and has ${\bf b}_4$; and client $4$ wants ${\bf b}_4$ and has ${\bf b}_3$. In this case, an optimal ({\it i.e.,} with the minimum number of transmissions) index code consists of sending $2$ transmissions, namely ${\bf b}_1 + {\bf b}_2$ and ${\bf b}_3 + {\bf b}_4$: it is easy to see that each client can decode the requested message from one of these transmissions using the side information. 
However, this index code can allow curious clients to violate the privacy of other clients who share the broadcast channel, by learning information that pertains to their requests and/or side information sets. For example, assume that client $4$ is curious. Upon learning the two transmissions, client $4$ knows that nobody is requesting message ${\bf b}_5$. Moreover, she knows that if a client is requesting ${\bf b}_1$ or ${\bf b}_2$ (similarly, ${\bf b}_3$ or ${\bf b}_4$), then this client should have the other message as side information in order to decode the requested message.

The solution {that we propose to limit this privacy leakage} stems from the following observation: it may not be necessary to provide clients with the entire set of broadcast transmissions. Instead, each client can be given access, and learn the coding operations, for only a subset of the transmissions, {\it i.e.,} the subset that would allow her to decode the message that she requested. Consider again the example in Figure~\ref{fig::IndexCodingExample}. The optimal index code consists of  two transmissions. However, each client is able to decode her request using exactly one of the two transmissions. Therefore, if each client only learns the coding coefficients for the transmission that she needs, then she will have no knowledge of the content of the other transmission, and  thus would have less information about the requests 
of the other clients. Limiting the access of each client to just one  out of the two transmissions was possible for this particular example; however, it is not the case that every index code  has this property.

Our approach in this paper builds on {the idea described above. In particular, {given an index coding instance that uses $T$ transmissions,} we ask:} Can we limit the access of each client to {at most $k \leq T$ transmissions,} while still allowing each client to decode {her} requested message? In other words, for a given index coding instance, what is the best (in terms of number of transmissions) index code that we can design such that each client is able to decode {her} request using at most $k$ out of these transmissions? Our work attempts to understand the fundamental relation between limiting the accessibility of clients to the coding matrix and the attained level of privacy.
In particular, we propose the use of {\it $k$-limited-access schemes}, that transform the coding matrix so {as to} restrict each client to access at most $k$ rows of the transformed matrix, as opposed to the whole of it. 
Our contributions include:
\begin{itemize}
\item We formalize the intuition that using $k$-limited-access-schemes can indeed increase the attained level of privacy against curious clients. We demonstrate this using two privacy metrics, namely an {\it entropy-based} metric and the {\it maximal information leakage}. 
{In both cases,} we show that the attained level of privacy is linearly dependent on the value of $k$, {\it i.e.,} privacy increases linearly with the number of rows of the coding matrix {that} we hide. 
\item We design polynomial time (in the number of clients) universal $k$-limited-access schemes ({\it i.e.}, that do not depend on the structure of the coding matrix), and require a simple matrix multiplication.
We prove that these schemes are {order-}optimal in some regimes, in particular when either $k$ or $n$ (the number of clients) is large.
Interestingly, when $k$ is larger than a threshold, these schemes enable to restrict the amount of access to half of the coding matrix with an overhead of exactly one additional transmission. This result indicates that some privacy-bandwidth trade-off points can be achieved with minimal overhead.
\item   
{We propose algorithms} that depend on the structure of the {coding matrix and show that, when $n$ and $k$ are both small, they provide improved performance with respect to the universal schemes mentioned above.} 
These schemes use a graph-theory representation of the problem, and {are optimal} for some special instances. 
\item We provide analytical and numerical performance evaluations of our schemes. We show how our proposed $k$-limited-access schemes provide a bandwidth-privacy trade-off, namely how much bandwidth usage ({\it i.e.}, number of transmissions) is needed to achieve a certain level of privacy (captured by the value of $k$). We show that our proposed {schemes provide} a trade-off curve that is close to the lower bound when either $k$ or $n$ is {large.} In the case where both $n$ and $k$ are small, we show through numerical evaluations that our proposed algorithms give {an} average performance that is close to the lower bound.
\end{itemize}

The paper is organized as follows.
Section~\ref{sec::setup} introduces our notation, {formulates the} problem, and gives a geometric interpretation.
Section~\ref{sec::Privacy} discusses how $k$-limited-access schemes {limit the privacy leakage.}
Section~\ref{sec:MainRes} shows {the construction of} $k$-limited-access schemes {and proves their order-optimality when either $n$ or $k$ is large.
Section~\ref{sec::BipartiteGraphRep} designs algorithms which are better-suited for cases when both $n$ and $k$ are {small.
Section~\ref{sec::relatedWork}} discusses related work and Section~\ref{sec::conclusion} concludes the paper.
Some of the proofs are delegated to the appendices.

\section{Notation, Problem Formulation and Geometric Interpretation}
\label{sec::setup}

\smallskip
\noindent\textbf{Notation.} 
Calligraphic letters indicate sets;
$|\Set{X}|$ is the cardinality of $\Set{X}$;
$[n]$ is the set of integers $\{1,\cdots,n\}$;
boldface lower case letters denote vectors and boldface upper case letters indicate matrices;
given a vector $\mathbf b$, $b_i$ indicates the $i$-th element of $\mathbf b$;
given matrices $\mathbf{A}$ and $\mathbf{B}$, $\mathbf{B} \subset_{k} \mathbf{A}$ indicates that $\mathbf B$ is formed by a set of $k$ rows of $\mathbf A$;
$\mathbf{0}_{j}$ is the all-zero row vector of dimension $j$;
$\mathbf{1}_{j}$ denotes a row vector of dimension $j$ of all ones and $\mathbf{I}_{j}$ is the identity matrix of dimension $j$;
$\mathbf{e}^j_i$ is the all-zero row vector of length $j$ with a $1$ in position $i$;
for all $x \in \mathbb{R}$, the floor and ceiling functions are denoted with $\lfloor x \rfloor$ and $\lceil x \rceil$, respectively;
logarithms are in base 2;
$\mathbf{Pr}(X)$ refers to the probability of event $X$.

\smallskip
\noindent\textbf{Index Coding.} We consider an index coding instance, where a server has a database $\mathcal{B}$ of $m$ messages $\mathcal{B}=\left \{ \mathbf{b}_{\Set{M}} \right \}$, where $\Set{M} = [m]$ is the set of message indices, and $\mathbf{b}_j \in \mathbb{F}_{2}^{F}, j \in \mathcal{M},$ with $F$ being the message size, and where operations are done over the binary field.
The server is connected through a broadcast channel to a set of clients $\mathcal{C}=\left \{ c_{\Set{N}} \right \}$, where $\Set{N} = [n]$ is the set of client indices. 
We assume that $m \geq n$.
Each client {$c_i, i \in \mathcal{N},$} has a subset of the messages {$\left \{ \mathbf{b}_{\Set{S}_i}\right \}$, with $\Set{S}_i \subset \Set{M}$,} as side information and requests a new message {$\mathbf{b}_{q_i}$} with $q_i \in \Set{M} \setminus \Set{S}_i$ that she does not have. 
We assume that the server employs a \textit{linear code}, {\it i.e.}, it designs a set of broadcast transmissions that are linear combinations of the messages in {$\mathcal{B}$}.  The linear index code can be represented as 
{$\mathbf{A} \mathbf{B} = \mathbf{Y}$,}
where {$\mathbf{A} \in \mathbb{F}_2^{T \times m}$} is the coding matrix, $\mathbf{B} \in \mathbb{F}_2^{m \times F}$ is the matrix of all the messages and $\mathbf{Y} \in \mathbb{F}_2^{T \times F}$ is the {resulting matrix} of linear combinations.
Upon receiving {$\mathbf{Y}$,} client $c_i{, i \in \mathcal{N},}$ employs linear decoding
 to decode the requested message {$\mathbf{b}_{q_i}$.}
\smallskip

\noindent\textbf{Problem Formulation.}
In~\cite{bar2011index}, it was shown that the index coding problem is equivalent to the rank minimization of an $n \times m$ matrix  $\mathbf{G} \in \mathbb{F}_2^{n \times m}$, whose $i$-th row $\mathbf{g}_i$, $i \in [n],$ has the following properties: (i) has a $1$ in the position $q_i$ ({\it i.e.,} the index of the message requested by client $c_i$), (ii) has a $0$ in the $j$-th position for all $j \in  \Set{M} \setminus \Set{S}_i$, (iii) can have either $0$ or $1$ in all the remaining positions. 
For instance, with reference to the example in Figure~\ref{fig::IndexCodingExample}, we would have
\begin{align*}
\mathbf{G} = 
\begin{bmatrix}
1 & \star & 0 & 0 & 0 \\
\star & 1 & 0 & 0 & 0 \\
0 & 0 & 1 & \star & 0 \\
0 & 0 & \star & 1 & 0 
\end{bmatrix},
\end{align*}
where $\star$ can be either $0$ or $1$.
It was shown in~\cite{bar2011index} that
finding an optimal linear coding scheme {\it i.e.,} with minimum number of transmissions) is equivalent to completing $\mathbf{G}$ ({\it i.e.,} assign values to the $\star$ components of $\mathbf{G}$) so that it has the minimum possible rank. 
Once we have completed $\mathbf{G}$, we can use a basis of the row space of $\mathbf{G}$ (of size $T = \text{rank} \left (\mathbf{G} \right)$) as a coding matrix  $\mathbf{A}$. 
In this case, client $c_i$ can construct $\mathbf{g}_i$ as a linear combination of the rows of $\mathbf{A}$, {\it i.e.,} $c_i$ performs the decoding operation $\mathbf{d}_i \mathbf{A} \mathbf{B} = \mathbf{d}_i \mathbf{Y}$, where $\mathbf{d}_i \in \mathbb{F}_2^T$ is the decoding row vector of $c_i$ chosen such that $\mathbf{d}_i \mathbf{A} = \mathbf{g}_i$.
{Finally, client $c_i$ can successfully decode $\mathbf{b}_{q_i}$ by subtracting from $\mathbf{d}_i \mathbf{Y}$ the messages corresponding to the non-zero entries of $\mathbf{g}_i$ (other than the requested message).
We remark that any linear index code that satisfies all clients with $T$ transmissions (where $T$ is not necessarily optimal) -- and can be obtained by any index code design algorithm~\cite{esfahanizadeh2014matrix,huang2015index,chaudhry2008efficient} -- 
corresponds to a completion of $\mathbf{G}$
(i.e., given $\mathbf{A} \in \mathbb{F}_2^{T \times m}$, we can create a corresponding $\mathbf{G}$ in polynomial time).


In our problem formulation we assume that we start with a given matrix $\mathbf{G}$ of rank $T$, {\it i.e.,}
we are given $n$ {\it distinct} vectors that belong to a $T$-dimensional subspace.
Using a basis of the row space of the given $\mathbf{G}$, we construct $\mathbf{A} \in \mathbb{F}_2^{T\times m}$.
Then, we ask:
\textit{Given $n$ distinct vectors  $\mathbf{g}_i$, $i \in [n]$, in a $T$-dimensional space, can we find a minimum-size set $\mathcal{A}_k$ with $T_k\geq T$ vectors, such that each $\mathbf{g}_i$ can be expressed as a linear combination of at most $k$ vectors in $\mathcal{A}_k$ (with $1 \leq k \leq T$)?}
The vectors in  $\mathcal{A}_k$ form the rows of the coding matrix $\mathbf{A}_{k}$ that we will employ. Then by definition, client $c_i$ will be able to reconstruct ${\mathbf g}_i$ using the matrix ${\mathbf A}_k^{(i)} \subset_k \mathbf{A}_k$.
We can equivalently restate the question as follows:
\textit{Given a coding matrix  $\mathbf{A}$,
can we find $\mathbf{P} \in \mathbb{F}_2^{T_k \times T}$, with $T_k$ as small as possible, such that
${\mathbf{A}_{k}} = \mathbf{P} \mathbf{A}$
and each row of $\mathbf{G}$ can be reconstructed by combining at most $k$ rows of $\mathbf{A}_{k}$?}                                                
Note that $k = T$ corresponds to the conventional transmission scheme of an index coding problem for which $\mathbf{P} = {\mathbf{I}_{T}}$.
In the remainder of the paper we will refer to a scheme that {\karmoose chooses} $\mathbf{A}_{k}$ {\karmoose to be the} coding matrix as $k$-limited-access scheme.

\smallskip

\begin{figure}
 \centering
   \subfigure[Conventional Transmission Protocol.]{
 \includegraphics[width=2.45in]{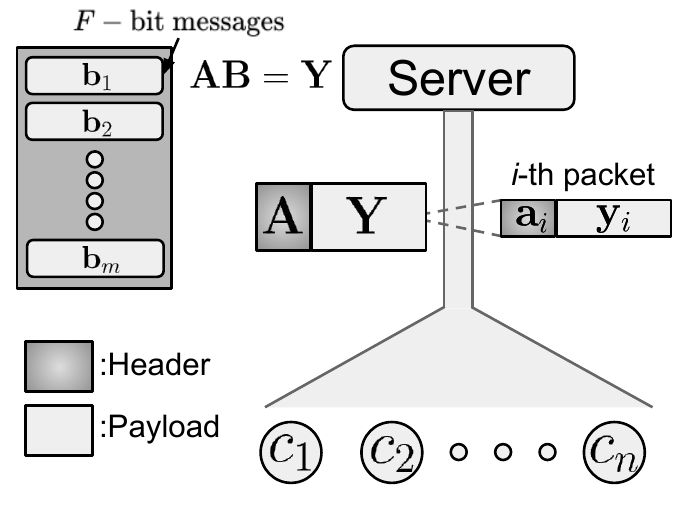}
\label{fig::ConvTransProt}
}
  \subfigure[Proposed Transmission Protocol.]{
 \includegraphics[width=2.45in]{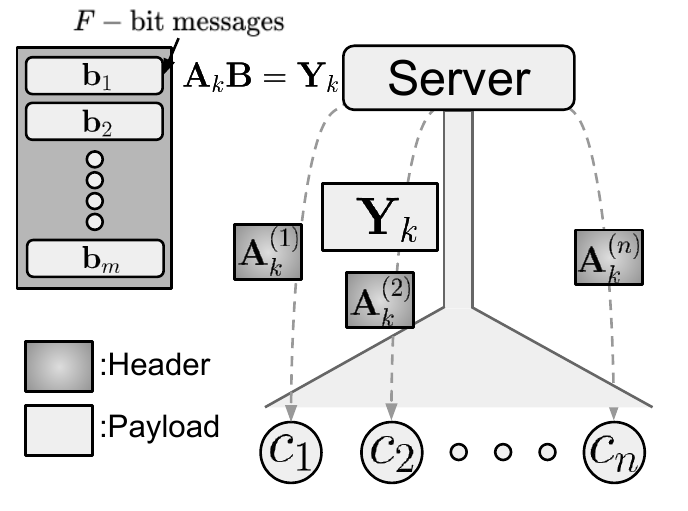}
\label{fig::NewTransProt}
}
\caption{A comparison between the conventional and the proposed transmission protocols. The proposed transmission protocol incurs a negligible increase in the transmission overhead when both $n$ and $m$ are $o(F)$.}
\label{fig::TransProt}
\end{figure}

\noindent\textbf{Transmission Protocol.}
In order to realize the privacy benefits of using $k$-limited-access schemes {-- which we will thoroughly illustrate in Section~\ref{sec::Privacy} -- we propose} a different transmission protocol for the index coding setup. Figure~\ref{fig::TransProt} shows both {the conventional and the proposed} transmission protocols. In the conventional protocol, {the} server designs a set of $T$ packets, each corresponding to an equation from the set of equations $\mathbf{AB} = \mathbf{Y}$. As shown in Figure~\ref{fig::ConvTransProt}, packet $i \in [T]$ consists of {(i)} a payload which contains the {linear} combination ${\mathbf{y}_i}$ and {(ii)} a header which contains the coefficients ${\mathbf a}_i$ used to create the equation. In the conventional protocol, {the} server sends {these} packets (both headers and payloads) on the broadcast channel to all clients. Our proposed protocol, however, {operates} differently. Specifically, {the} server {generates} packets which {correspond} to the set of equations ${\mathbf A}_k \mathbf{B} = \mathbf{Y}_k$ in a way that is similar to the conventional protocol. The server then {sends} {\it only} the payloads of these packets on the broadcast channel. Differently, the server {sends} the coefficients corresponding to {\it only} {$\mathbf{A}_k^{(i)}\subset_k \mathbf{A}_k$} to client $c_i$ using a private key or on a dedicated private channel ({\it e.g.,} the same channel used by $c_i$ to convey {her} request to the server).
Thus, using a $k$-limited-access scheme incurs an extra transmission overhead to privately convey the coding vectors.
In particular, the total number of transmitted bits $\text{C}_k$ {can be} upper bounded {as}
$\text{C}_k \leq n k m + T_k F,$
while the total number of transmitted bits 
$\text{C}$ using a conventional scheme is $\text{C} = T (F+m)$. 
{The extra} overhead incurred  is negligible in comparison to the broadcast transmissions that convey the encoded messages when $n$ and $m$ are both $o(F)$, which is a reasonable assumption for large file sizes (for instance, when sharing YouTube videos).

\begin{figure}
 \centering
 \includegraphics[width=0.65\columnwidth]{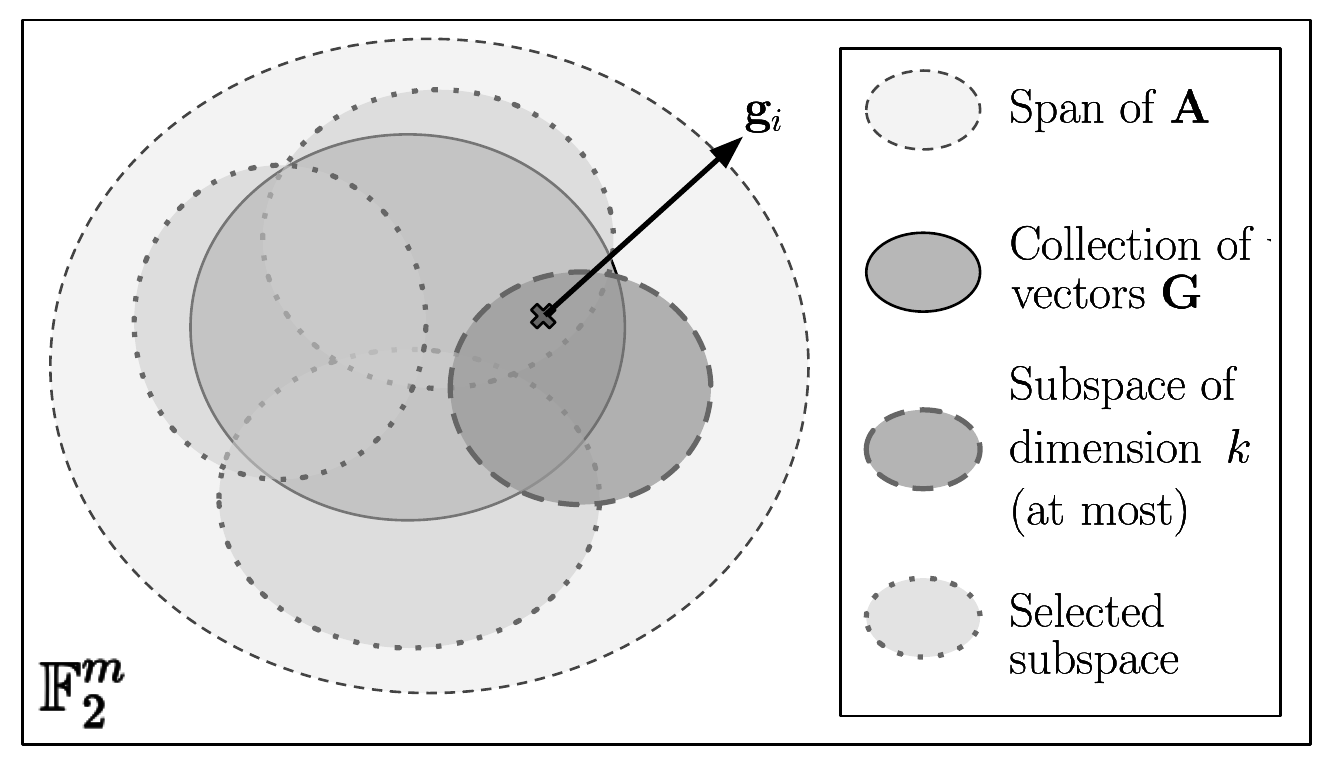}
 \caption{A geometric interpretation of $k$-limited-access schemes. An index code $\mathbf{A}$ is obtained from a particular filling of the matrix $\mathbf{G}$. Therefore, the collection of row vectors of $\mathbf{G}$ lies in the span of $\mathbf{A}$. Finding $\mathbf{A}_k$ is equivalent to finding a collection of subspaces, each of dimension at most $k$, to cover $\mathbf{G}$. Client $c_i$ is sent a collection of (at most) $k$ rows of $\mathbf{A}_k$; these correspond to one subspace which covers $\mathbf{g}_i$.  }
 \label{fig::GeomInterp}
\end{figure}

\smallskip

\noindent\textbf{Geometric Interpretation.}
The geometric interpretation of our problem is depicted in Figure~\ref{fig::GeomInterp}. An index code $\mathbf{A}$ corresponds to a particular completion of the matrix $\mathbf{G}$. Therefore, the set of row vectors in $\mathbf{G}$ lies in the row span of $\mathbf{A}$ (which is of dimension $T$). We denote this subspace of dimension $T$ by $L$. The problem of finding a matrix $\mathbf{A}_k$ can be interpreted as finding a set of subspaces, each of dimension at most $k$, such that each row vector $\mathbf{g}_i$, $i \in [n]$, is covered by at least one of these smaller subspaces. Once these subspaces are selected, then the rows of $\mathbf{A}_k$ are taken as the union of the basis vectors of all these subspaces. Client $c_i$ is then given the basis vectors of subspace $L_i$, {\it i.e.}, the one which covers $\mathbf{g}_i$, instead of the whole matrix $\mathbf{A}_k$. Therefore $c_i$ would have perfect knowledge of $L_i$ instead of $L$. Having less information about $L$  naturally translates to less information about the requests of other clients, as we more formally discuss in the next section.

\section{Achieved Privacy Levels}
\label{sec::Privacy}

In this section, we investigate and quantify the level of privacy that
$k$-limited-access schemes
can achieve compared to a conventional index coding scheme ({\it i.e.,} when each client has access to the entire coding matrix).
In what follows, we consider the setup {described} in the previous section and suppose that client $c_n$ is curious, {\it i.e.,} by leveraging the {(at most)} $k$ rows $\mathbf{A}^{(n)}_k$ that she receives, she seeks to infer information about client $c_i, i \in [n-1]$.
{Specifically,} we are interested in quantifying the amount of information that $c_n$ can obtain about $q_i$ ({\it i.e.,} the identity of the request of $c_i$) as a function of $k$.

We assume that the index coding instance is random, {\it i.e.}, we consider the requests and side information sets of clients as random variables and denote them as $Q_{[n]}$ and $S_{[n]}$, respectively.
The operation of the server is shown in Figure \ref{fig::kflow} and is described as follows:

 \noindent {\bf{Step-1:}} The server obtains the information about the requests $Q_{[n]}$ and side information sets $S_{[n]}$ of all clients $c_{[n]}$.

  \noindent {\bf{Step-2:}} Based on this information, the server designs an index code $\mathbf A$ {by means of} some {index coding} algorithm~\cite{esfahanizadeh2014matrix,huang2015index,chaudhry2008efficient}.
 
\noindent {\bf{Step-3:}} The server then applies the $k$-limited-access scheme to obtain $\mathbf{A}_k = \mathbf{P} \mathbf{A}$, {where $\mathbf{P}$ is a deterministic mapping from $ \mathbf{A}$ to $\mathbf{A}_k$ (see Section~\ref{sec:MainRes} for the construction of $\mathbf{P}$). This implies that $T_k$ is a deterministic function of $T$ {and $k$ ({\it i.e.,} the parameter of the scheme)}.}
 
\noindent {\bf{Step-4:}} The server sends $\mathbf{A}_k^{(i)}$ to client $c_i$. 
{If multiple $\mathbf{A}_k^{(i)}$ can be selected, then}
the server picks {and transmits one such matrix}
uniformly at random, {independently} of the underlying $\mathbf{A}$ which might have generated this $\mathbf{A}_k$.

{We are now interested in quantifying the level of privacy that is achieved by the protocol described above. Towards this end,} we use two privacy metrics, namely an {\it entropy-based} metric and the {\it maximal information leakage}.

\begin{figure}
 \centering
 \includegraphics[width=6.5in]{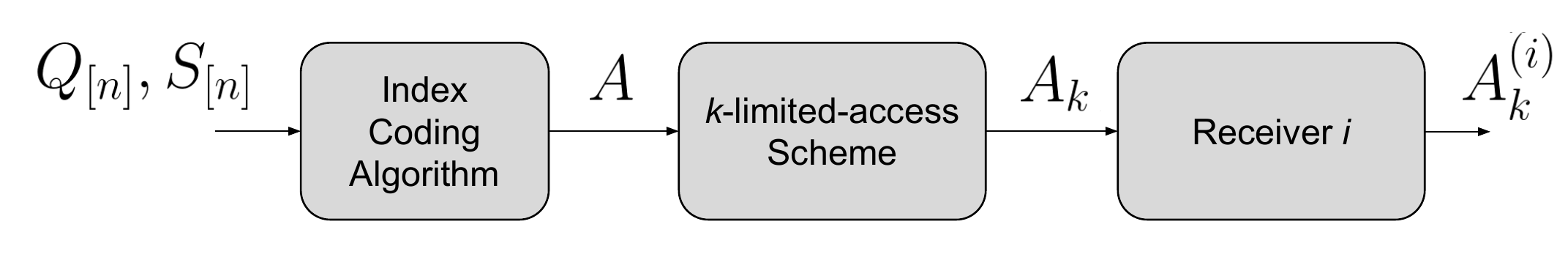}
 \caption{The procedure of designing an index code and applying $k$-limited-access schemes.}
 \label{fig::kflow}
\end{figure}

\subsection{Entropy-Based {Privacy Metric}}


The entropy-based privacy metric is inspired by the geometric interpretation of our problem in Figure~\ref{fig::GeomInterp}.
We let $L$ (respectively, $L_n$) be the random variable associated with the subspace spanned by the $T$ rows of the  coding matrix $\mathbf{A}$ (respectively, spanned by the $k$ {row} vectors of $\mathbf{A}_k^{(n)}$).
Client $c_n$ receives the matrix $\mathbf{Y}_k$ and as such she knows $T_k$. 
Given this, we now define the entropy-based privacy metric and evaluate it for the proposed protocol.
%
\begin{definition}
The entropy-based privacy metric is defined as \[P_k^{(\text{Ent})} = H \left(L|L_n, T_k\right),\] and quantifies the amount of uncertainty  that $c_n$ has about the subspace spanned by the $T$ rows of the index coding matrix $\mathbf{A}$.
\end{definition}

Before characterizing $P^{(\text{Ent})}_k$, we state the following lemma, which is proved in Appendix~\ref{app::subsetlen}.
\begin{lemma}
 \label{lem::subspaceCount}
 Given a subspace $L_n \subseteq \mathbb{F}_2^{m}$ of dimension $k$, let $\mathcal{L} (T, L_n)$ be the set of subspaces $L \subseteq \mathbb{F}_2^{m}$ of dimension $T \geq k$ where $L_n \subseteq L$. Then $|\mathcal{L} (T, L_n)|$ is equal to
 \begin{equation*}
 |\mathcal{L} (T, L_n)| = \prod_{\ell=0}^{T-k-1} \frac{2^m-2^{k+\ell}}{2^T-2^{k+\ell}}.
 \end{equation*}
\end{lemma}

Assume {an index} coding setting with $c_n$ observing a particular subspace $L_n = \ell_n$ and a number of transmissions {$T_k = t_k$ for the $k$-limited access scheme.} {\karmoose Moreover, we consider a stronger adversary ({\it i.e.}, curious client) and assume that she also knows the specific realization of $T = t$.}
Given this, we can compute 
\begin{align}
P^{(\text{Ent})}_k &=  {\karmoose H }\left( L | L_n = \ell_n, T_k=t_k, T = t  \right ) \nonumber \\
& \stackrel{{\rm{(a)}}}{=} H \left( L | L_n = \ell_n, T=t  \right ) \stackrel{{\rm{(b)}}}{=} \log \left( |\mathcal{L} (t, \ell_n)|\right) \nonumber
\\& \stackrel{{\rm{(c)}}}{=}\log \left(  \prod_{\ell=0}^{t-k-1} \frac{2^m-2^{k+\ell}}{2^t-2^{k+\ell}} \right ) \stackrel{m \gg t}{\approx} m(t-k),
\label{eq:ent}
\end{align}
where:
{\karmoose (i)} the equality in {\karmoose $\rm{(a)}$} follows because $T_k$ is a deterministic function of $T$ and $k$, which is the parameter of the scheme (see Step-3);
{\karmoose (ii)} the equality in $\karmoose \rm{(b)}$ follows by assuming that the underlying system maintains a uniform distribution across all feasible $t$-dimensional subspaces of $\mathbb{F}_2^m$;
{\karmoose (iii)} the equality in ${\karmoose \rm{(c)}}$ follows by virtue of Lemma~\ref{lem::subspaceCount}.  
%
{\karmoose We} note that when $m \gg t$, then the {\karmoose quantity}
in~\eqref{eq:ent} decreases linearly with $k$, {\it i.e.,} as intuitively {expected,} the less rows of the coding matrix $c_n$ learns, the less she can infer about the subspace spanned by the $T$ rows of the coding matrix $\mathbf{A}$.
This suggests that, by increasing $k$, $c_n$ has {less} uncertainty about $q_i$.
{Note also that $P_k^{(\text{Ent})}$ is zero} when $k=t$; this is because, under this condition, $c_n$ receives the entire index coding matrix, {\it i.e.}, $L_n = L$,  and hence she is able to perfectly reconstruct the subspace spanned by its rows.
However, although $P_k^{(\text{Ent})} = 0$ when $k=t$, {\karmoose $c_n$} might still have uncertainty about $q_i$~\cite{karmoose2017private}.
Quantifying this uncertainty is an interesting open problem; this uncertainty, in fact, depends on the underlying system, {\it e.g.,} on the index code used by the server and on the distribution with which the index code matrix is selected.


\subsection{Maximal Information Leakage}

The second metric {that} we consider as our privacy metric is the Maximal Information Leakage (MIL)~\cite{issa2016operational}. 
{Given two discrete random variables $X$ and $Y$ with alphabets $\mathcal{X}$ and $\mathcal{Y}$, the MIL from $X$ to $Y$ is denoted by $\mathcal{L}(X \rightarrow Y)$ and defined as}
\begin{align}
\label{eq::MILdef}
 \mathcal{L}(X \rightarrow Y) = 
{\sup\limits_{S-X-Y } \log  \frac{\sum_{y \in \mathcal{Y}} \max_{s \in \mathcal{S}} p_{SY}(s,y)}{\max_{s \in \mathcal{S}}p_S(s)}} 
 =\log \sum\limits_{y \in \mathcal{Y}} \max\limits_{x \in \mathcal{X} : p_X(x) > 0} p_{Y|X}(y|x),
\end{align}
where the second equality
is shown in~\cite{issa2016operational}. The MIL metric captures the amount of information leaked about $X$ through $Y$ to an adversary, who is interested in estimating a (possibly probabilistic) function $S$ of $X$. This is captured by the fact that $S - X - Y$ forms a Markov chain as shown in {the expression in~\eqref{eq::MILdef}.} The metric considers a worst-case such adversary, that is, an adversary who is interested in computing a function $S$ for which the maximum information can be leaked out of $Y$. 
The result in~\cite{issa2016operational} shows that this quantity depends only on the joint distribution of $X$ and $Y$. The following properties of the MIL are useful~\cite{issa2016operational}:
\begin{itemize}
 \item (Property 1): If $X-Y-Z$, then $\mathcal{L}(X\rightarrow Z) \leq \min\{ \mathcal{L}(X \rightarrow Y), \mathcal{L}(Y \rightarrow Z) \}$,
 \item (Property 2): $\mathcal{L}(X \rightarrow Y) \leq \min\{ \log |\mathcal{X}|, \log |\mathcal{Y}|\}$,
 \item (Property 3): $\mathcal{L}(X \rightarrow X) = \log |\left\{x : p_X(x) > 0 \right\}|$.
\end{itemize}
To describe how we use the MIL as a privacy metric in our setup, we  first need to define what are the corresponding random variables $X$ and $Y$, and then argue that the estimation of client $c_n$ of the requests of other clients  forms a Markov chain as required by the MIL definition. To do so, we first define the following sets:

\noindent\textit{1)} Given $\mathbf{g}_i$, $\mathbf{A}_k$ and an integer $r$,  let $\mathcal{P}(\mathbf{g}_i,\mathbf{A}_k, r)$ be the set of all possible sub-matrices $\mathbf{A}_k^{(i)}$ of $\mathbf{A}_k$ with {\it exactly} $r$ rows, that client $c_i$ can use to reconstruct the vector $\mathbf{g}_i$:
\begin{align*}
 \mathcal{P}(\mathbf{g}_i,\mathbf{A}_k, r) = \left\{ \mathbf{Z} \subset_r \mathbf{A}_k \: | \:  \exists \mathbf{d} \in \mathbb{F}_2^r \text{ s.t. } \mathbf{g}_i = \mathbf{d} \mathbf{Z} \right\}, \nonumber
\end{align*}

\noindent\textit{2)} Given $q_i$, $\Set{S}_i$ and $\mathbf{A}_k$, let $\mathcal{T}(q_i,\mathcal{S}_i,\mathbf{A}_k)$ be the set of all possible sub-matrices $\mathbf{A}_k^{(i)}$ of $\mathbf{A}_k$ with the minimum possible number of rows, such that client $c_i$ with side information $\mathcal{S}_i$ can decode~$q_i$:
\begin{equation*}
  \mathcal{T}(q_i,\mathcal{S}_i,\mathbf{A}_k) = \bigcup\limits_{\mathbf{g}_i \in \Set{G}(q_i, \mathcal{S}_i)} \mathcal{P}(\mathbf{g}_i,\mathbf{A}_k, r_{\text{min}}),
\end{equation*}
where
\begin{equation}
 \mathcal{G}(q_i,\mathcal{S}_i) = \left\{ \mathbf{g} \in \mathbb{F}_2^{m} \: | \:  g_{q_i} = 1, g_{[m] \setminus \{ q_i \cup \mathcal{S}_i \}} = 0 \right\}, \nonumber
\end{equation}
and
\begin{align*}
r_{\text{min}} = \min \mathcal{R}, \ \mathcal{R} = \left \{ r \in \mathbb{N}^{+} : \ \exists \mathbf{g}_i \in \Set{G}(q_i, \Set{S}_i) \text{ such that }  \mathcal{P}(\mathbf{g}_i,\mathbf{A}_k, r) \neq \emptyset  \right \}.
\end{align*}

Since the requests and the side information sets are considered as random variables, then all subsequently generated codes, namely $\mathbf{A}$, $\mathbf{A}_k$ and $\mathbf{A}_k^{(i)}$ can be treated as random variables as well. We denote the corresponding random variables of these quantities as $A$, $A_k$ and $A_k^{(i)}$ respectively. In other words, for a given realization of $Q_{[n]} = q_{[n]}$ and $S_{[n]} = \Set{S}_{[n]}$, the corresponding realizations of the aforementioned codes used by the server are $A = \mathbf{A}$, $A_k = \mathbf{A}_k$ and $A_k^{(i)} = \mathbf{A}_k^{(i)}$.

When using conventional index codes ({\it i.e.,} without $k$-limited-access schemes), client $c_n$ ({\it i.e.,} the curious client and hence the adversary) would try to infer information about $Q_{[n-1]}$ from observing $A$ and given her information of $Q_n,S_n$. Therefore, one can think of client $c_n$ estimate of $Q_{[n-1]}$ as being a particular estimation function, the input of which is $A$. 
Differently, after using $k$-limited-access schemes, client $c_n$ would only have observed $A_k^{(n)}$ instead of $A$. 
Therefore, in the context of MIL, one choice of the variables $X$ and $Y$ is $A$ and $A_k^{(n)}$ respectively. The function $S$ would therefore be client $c_n$'s estimate of $Q_{[n-1]}$ out of $A$. The following proposition shows that this choice of variables $X$, $Y$ and $S$ allows us to use the MIL as a metric.


\begin{proposition}
 The following Markov chain holds
 \begin{align}
 \label{eq::markovchain}
 Q_{[n-1]} - A - A_k - A_k^{(n)},
\end{align}
conditioned on the knowledge of $Q_n,S_n$ in every stage of the chain.
\end{proposition}
\begin{Pf}
We have the following:
\begin{itemize}
 \item $Q_{[n-1]}  - A - A_k$ holds since $A_k$ is a deterministic function of $A$ (see also Step-3 of the proposed protocol);
 \item $A - A_k - A_k^{(n)}$ holds since $p(A_k^{(n)}| A_k,Q_n,S_n) = 1/|\mathcal{T}(Q_n,S_n,A_k) |$, independent of $A$, as described in Step-4 of the proposed protocol.
 \end{itemize}
\end{Pf}

\noindent We define $P_k^{(\text{MIL})} = \mathcal{L}\left( A \rightarrow A_k^{(n)} | Q_{n}=q_n ,S_{n}=\mathcal{S}_n \right)$ as our MIL privacy metric\footnote{We use the notation $\mathcal{L}\left(X \rightarrow Y | Z \right)$ to denote that the variables $X$ and $Y$ are conditioned on $Z$.}. The quantity $P_k^{(\text{MIL})}$ gives the maximum amount of information that $c_n$ can extract about $Q_{[n-1]}$ given the knowledge of $Q_{n},S_{n}$. 
The following theorem -- proved in Appendix~\ref{app::P_MIL} -- provides a guarantee on $P_k^{(\text{MIL})}$.
%
\begin{theorem}
 \label{thm::P_MIL}
Using the MIL, the attained level of privacy against a curious client when $k$-limited-access schemes are used is
 \begin{equation}
 \label{eq::P_MIL}
  P_k^{(\text{MIL})} = O(|\Set{S}_n| + mk).
 \end{equation}
\end{theorem}
\noindent The quantity in~\eqref{eq::P_MIL} characterizes the maximum amount of information that can be leaked to a curious client when $k$-limited-access schemes are used. It is clear that decreasing $k$ would decrease this amount of information; this aligns with the intuition that the less rows a server gives to a client, the less information a client would be able to infer about other clients sharing the broadcast domain. In order to shed more light on the benefits of using $k$-limited-access schemes, one could compare the quantity $P_k^{(\text{MIL})}$ with the MIL obtained when $k$-limited-access schemes are not used, {\it i.e.,} when a client observes the whole matrix $A$. Let this quantity be denoted as $\bar{P}_k^{(\text{MIL})} = \mathcal{L}(A \rightarrow A | Q_n=q_n,S_n=\mathcal{S}_n)$. Then we have the following result, which is proved in Appendix~\ref{app::P_MIL_Conv}.

\begin{theorem}
\label{thm::P_MIL_Conv}
Using the MIL, the attained level of privacy against a curious client for a conventional index coding setup is
\begin{equation}
 \label{eq::P_MIL_Conv}
 \bar{P}_k^{(\text{MIL})} = \Omega \left( mT - T^2 \right).
\end{equation}
\end{theorem}

\begin{figure}
 \centering
 \includegraphics[width=6in]{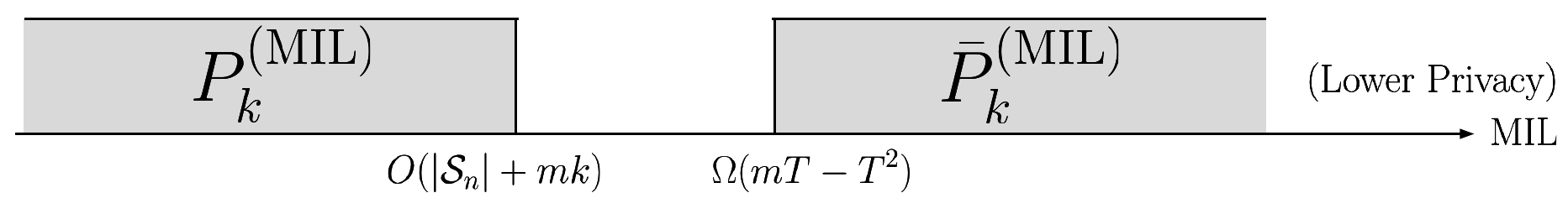}
 \caption{This figure shows how the MIL privacy metrics compare for the conventional index coding schemes and the $k$-limited-access schemes. Taking $k = o(T)$ would guarantee privacy gains when using  $k$-limited-access schemes.}
 \label{fig::MILComp}
\end{figure}

\noindent The results in Theorem~\ref{thm::P_MIL} and Theorem~\ref{thm::P_MIL_Conv} can be interpreted with the help of Figure~\ref{fig::MILComp}. The $k$-limited-access schemes achieve privacy gains as compared to conventional index codes, when the two bounds in~\eqref{eq::P_MIL} and~\eqref{eq::P_MIL_Conv}  strictly mismatch. A sufficient (but not necessary) condition for this is to select $k = o(T)$.

%

\section{Construction of $k$-limited-access Schemes}
\label{sec:MainRes}
In this section, we focus on designing $k$-limited-access schemes and assessing their theoretical performance in terms of number of additional transmissions required with respect to a {conventional} index coding scheme.
Recall that we are given a coding matrix  $\mathbf{A}$ that requires $T$ transmissions.
Then, we seek to construct a matrix  $\mathbf{P} \in \mathbb{F}_2^{T_k \times T}$, so that $\mathbf{A}_k = \mathbf{P} \mathbf{A}$, and each client needs to access at most $k$ rows of $\mathbf{A}_k$ to decode her requested message. In particular, we aim at constructing
matrices  $\mathbf{P}$ with $T_k$ as small as possible.
Trivially, $T_k\geq T$.
Towards this end, we first derive upper and lower bounds on $T_k$. 
Our main result is stated in the theorem below.
\begin{theorem}
\label{theorem_main}
Given an index coding matrix $\mathbf{A} \in \mathbb{F}_2^{T \times m}$ with $T \geq 2$, it is possible to transform it into ${\mathbf{A}_{k}} = \mathbf{P} \mathbf{A}$ with $\mathbf{P} \in \mathbb{F}_2^{T_k \times T}$, such that each client can decode her requested message by combining at most $k$ rows of ${\mathbf{A}_{k}}$, if and only if
\begin{align}
 T_k \geq \max \left \{ T, T^\star \right \}, \quad T^\star= \min \left\{T_k : \sum\limits_{i=1}^{k} {T_k \choose i} \geq n \right\}. \label{eq::lb}
\end{align}
Moreover, we provide polynomial time (in $n$) constructions of $\mathbf{P}$ such that:
\begin{itemize}
\item When $ \lceil T/2 \rceil \leq k < T$, then 
\begin{align}
T_k  \leq \min \left\{n, T + 1 \right\};  \label{theorem_ub2}
  \end{align}
\item When $ 1 \leq k < \lceil T/2 \rceil$, then
\begin{align}
 T_k \leq \min \left\{n,k2^{\left\lceil \frac{T}{k} \right\rceil} \right\}.  \label{theorem_ub1}
\end{align}
\end{itemize}
 \end{theorem}

\begin{Pf}
{The lower bound on $T_k$ in~\eqref{eq::lb} is proved in Appendix~\ref{app::theorem_main}.}
In particular, the  bound in (\ref{eq::lb}) says that, if we are allowed to combine at most $k$ out of the $T_k$ vectors, then we should be able to create a sufficient number of vectors.
The two upper bounds on $T_k$ in~\eqref{theorem_ub2} and~\eqref{theorem_ub1} are proved in Section~\ref{eq:ConstrUB}, where we give explicit constructions for $\mathbf{P}$.
\end{Pf}

We note that, as expected, the smaller the value of $k$ that we require, the larger the value of $T_k$ that we need to use.
Trivially, for $k=1$ we would need $T_k=n$, {\it i.e.}, the server would need to send uncoded transmissions.
Thus, there is a trade-off
between the bandwidth -- measured as the number $T_k$ of broadcast transmissions -- and privacy -- captured by the value of $k$ that we require. 
Interestingly, when $k \geq \lceil T/2 \rceil $, with just one extra transmission, {\it i.e.}, $T_k=T+1$, we can  restrict the access of each client to at most half of the coding matrix, independently of the coding matrix $\mathbf{A}$.
In other words, for this regime, we can achieve a certain level of privacy with minimal overhead. However, as we further reduce the value of $k$, the overhead becomes more significant. 
%
%
%
Moreover, the results in Theorem~\ref{theorem_main} also imply that our constructions are order-optimal in the case of large values of $n$ (when $n = \Theta(2^T)$)\footnote{Note that $n$ is always $O(2^T)$ ({\it i.e.}, the number of distinct vectors $\mathbf{g}_i$ for a given $T$ is at most $2^T-1$). The case of large values of $n$ corresponds to the case where this bound on the number of distinct vectors $\mathbf{g}_i$ is not loose: there is a corresponding lower bound on $n$, {\it i.e.}, $n = \Omega(2^T)$. Therefore, the case of large values of $n$ corresponds to $n = \Theta(2^T)$.}. In addition, when $\lceil T/2 \rceil \leq k < T$, our scheme is at most one transmission away from the optimal number of transmissions, and this is for {\it any} value of $n$.
This is shown in the following lemma, which is proved in Appendix~\ref{app::theorem_main}.

\begin{lemma}
\label{lemma::SpecCase}
Consider an index coding setup.
We have
\begin{itemize}
\item When $n = 2^T-1$ and $ \lceil T/2 \rceil \leq k < T$, the bounds in~\eqref{eq::lb} and~\eqref{theorem_ub2} coincide, {\it i.e.}, the provided construction of $\mathbf{P}$ is optimal;
%
\item For any value of $n < 2^T -1$ and $ \lceil T/2 \rceil \leq k < T$, the bound in~\eqref{theorem_ub2} is at most one transmission away from the bound in~\eqref{eq::lb};
\item When $n = \Theta(2^T)$ and for any value of $k$, 
then $T_k = \Theta(k2^{\frac{T}{k}})$, {\it i.e.}, the provided construction is order-optimal.
\end{itemize}
\end{lemma}

\begin{figure*}
 \centering
 \subfigure[$n = 2^T-1$]{
  \centering
  \includegraphics[width = 0.31\columnwidth]{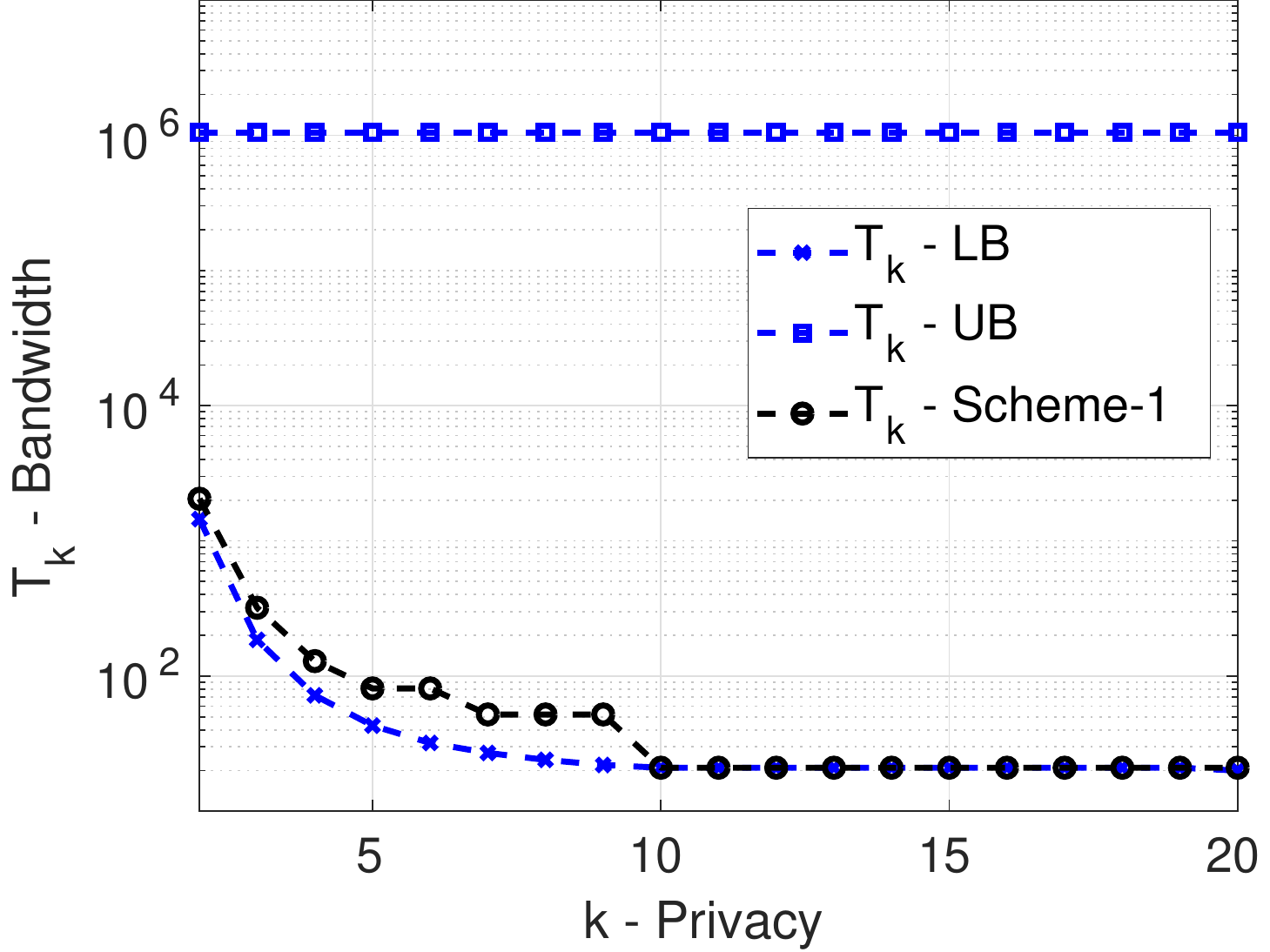}
  \label{fig::n_full}
 }
 \subfigure[$n = T^4$]{
  \centering
  \includegraphics[width = 0.31\columnwidth]{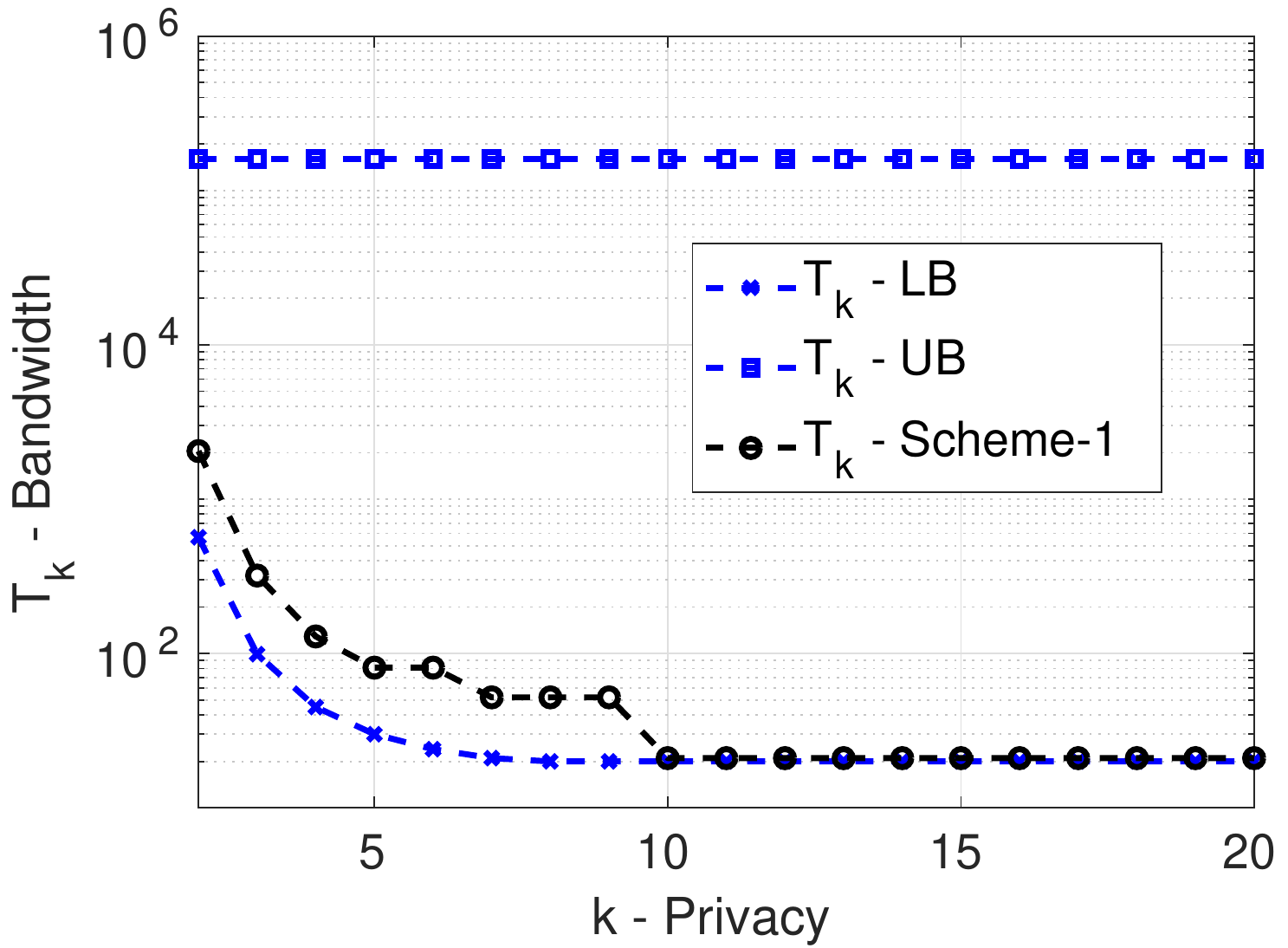}
  \label{fig::n_pow4}
 }
 \subfigure[$n = T^2$]{
  \centering
  \includegraphics[width = 0.31\columnwidth]{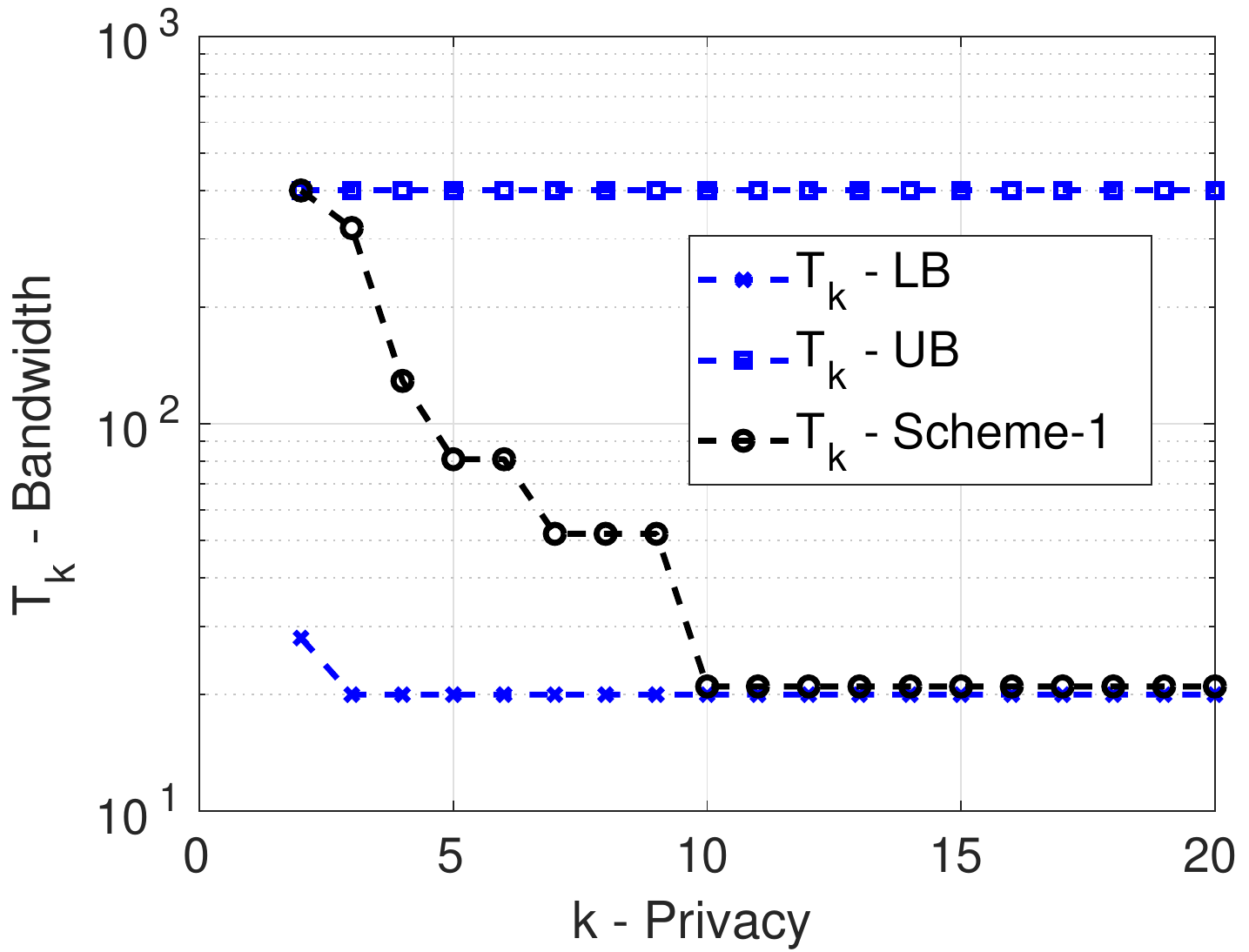}
  \label{fig::n_pow2}
 }
 \caption{Bandwidth ($T_k$ on the y-axis) versus privacy ($k$ on the x-axis) trade-off when using the $k$-limited-access schemes in Theorem~\ref{theorem_main} for different values of $n$. The plots in this figure are for $T = 20$.}
 \label{fig::BandwidthPrivacyTradeoff}
\end{figure*}

Figure~\ref{fig::BandwidthPrivacyTradeoff} shows the trade-off exhibited by our proposed $k$-limited-access schemes between bandwidth usage ($T_k$) and the attained privacy ($k$) - we use $k$ as a proxy to the amount of attained privacy against a curious client (see Section~\ref{sec::Privacy}). 
The figure shows the performance of our {constructions} in Theorem~\ref{theorem_main} (labeled as Scheme-1), as well as the lower bound {in~\eqref{eq::lb} (labeled as LB)} and an upper bound which corresponds to uncoded transmissions (labeled as UB). 
Figure~\ref{fig::n_full} confirms the order-optimality of our constructions when $n = 2^T-1$. In addition, our schemes perform similarly well when $n$ is sufficiently large (and not necessarily equal to $2^T-1$) as shown in Figure~\ref{fig::n_pow4} where $n = T^4$. 
Finally, Figure~\ref{fig::n_pow2} shows the performance for a small value of $n$ ($n = T^2$). The figure shows that our proposed {constructions do} not perform as well when $n$ and $k$ are small, a case which we study in more details in Section~\ref{sec::BipartiteGraphRep}.

We now conclude this section by giving explicit constructions of the $\mathbf{P}$ matrix {and prove the two upper bounds on $T_k$ in~\eqref{theorem_ub2} and~\eqref{theorem_ub1}.} 
Our design of $\mathbf{P}$ allows to reconstruct any of the $2^T$ vectors of size $T$.
As such our constructions are universal, in the sense that the matrix $\mathbf{P}$ that we construct does not depend on the specific index coding matrix $\mathbf{A}$.
%
%
%
 
\smallskip

\subsection{Proof of Theorem~\ref{theorem_main}, Equations~\eqref{theorem_ub2} and \eqref{theorem_ub1}}
\label{eq:ConstrUB}
Recall that $\mathbf{A}$ is full rank and that the $i$-th row of $\mathbf{G}$ can be expressed as $\mathbf{g}_i = \mathbf{d}_i \mathbf{A}$, where $\mathbf{d}_{i} \in \mathbb{F}_2^T$ is the coefficients {row} vector associated {with $\mathbf{g}_i$.} We next analyze two different cases/regimes, which depend on the value of $k$.
\noindent

\noindent\textbf{Case I:}
$ \lceil T/2 \rceil \leq k < T$.
When $n \geq T +1$, let
\begin{align}
\label{eq:PPartI}
 \mathbf{P} = \left[ \begin{matrix} \mathbf{I}_{T} \\ \mathbf{1}_T \end{matrix} \right],
\end{align}
\noindent which results in a matrix $\mathbf{A}_k$ with $T_k = T + 1$, {matching} {the bound} in~\eqref{theorem_ub2}.
%
We now show that each $\mathbf{g}_i = \mathbf{d}_i \mathbf{A}, {i \in [n],}$ can be reconstructed by combining up to $k$ vectors of $\mathbf{A}_k$. Let $w (\mathbf{d}_i)$ be the Hamming weight of $\mathbf{d}_i$. If $w(\mathbf{d}_i) \leq \lceil T/2 \rceil$, then we can reconstruct {$ \mathbf{g}_i$} as $\mathbf{g}_i = [\mathbf{d}_i \:\: 0] \mathbf{A}_k$, which involves adding $w(\mathbf{d}_i) \leq \lceil T/2 \rceil \leq k$ rows of $\mathbf{A}_k$. 
{Differently,} if $w(\mathbf{d}_i) \geq \lceil T/2 \rceil + 1$, then we can reconstruct $\mathbf{g}_i$ as ${\mathbf{g}_i} = [\bar{\mathbf{d}}_i \:\: 1] {\mathbf{A}_{k}}$, where $\bar{\mathbf{d}}_i$ is 
the bitwise complement of $\mathbf{d}_i$. 
In this case, reconstructing $\mathbf{g}_i$ involves adding $T - w(\mathbf{d}_i) + 1 \leq \lfloor T/2 \rfloor \leq k$ rows of $\mathbf{A}_k$. 

When $n < T+1$, then it is sufficient to send $n$ uncoded transmissions, where the $i$-th transmission satisfies $c_i, i \in [n]$. 
In this case $c_i$ has access only to the $i$-th transmission, {{\it i.e.,}} $k=1$.
This completes the proof of the upper bound in~\eqref{theorem_ub2}.

 \textbf{Example:} We show how the scheme works via a small example, where $T = 4$ and $k = 2$. In this case, we have 
 
 \begin{equation*}
\mathbf{P} = \left[ \begin{matrix}
1 & 0 & 0 & 0  \\
0 & 1 & 0 & 0 \\
0 & 0 & 1 & 0 \\
0 & 0 & 0 & 1 \\
1 & 1 & 1 & 1 \end{matrix}\right].
 \end{equation*}
 If $\mathbf{g}_i = \left[ \begin{matrix} 1 & 1 & 0 & 0 \end{matrix} \right]\mathbf{A}$, then it can be reconstructed as $\mathbf{g}_i = \left[ \begin{matrix} 1 & 1 & 0 & 0 & 0 \end{matrix} \right]\mathbf{PA}$ with $2$ rows of $\mathbf{PA}$ used in the reconstruction. Differently, if $\mathbf{g}_i = \left[ \begin{matrix} 1 & 1 & 1 & 0 \end{matrix} \right]\mathbf{A}$, then it can be reconstructed as $\mathbf{g}_i = \left[ \begin{matrix} 0 & 0 & 0 & 1 & 1 \end{matrix} \right]\mathbf{PA}$ with again $2$ rows of $\mathbf{PA}$ used in the reconstruction.\\


\noindent\textbf{Case II:}
$1 \leq k < \lceil T/2 \rceil$. 
Let $Q = \left\lfloor T/ \left\lceil \frac{T}{k} \right\rceil \right\rfloor$ and 
 $T_{\text{rem}} = T - Q\left\lceil \frac{T}{k}\right\rceil $. If $k$ divides $T$, then $Q = k$, $T_{\text{rem}} = 0$, otherwise $Q \leq k-1$ and $T_{\text{rem}} \leq \left\lceil \frac{T}{k} \right\rceil$. Then, we can write 
 
 \begin{equation*}
\mathbf{P} = \left[ \begin{matrix}
                      \mathbf{Z}_1 & \mathbf{0} & \cdots & \mathbf{0} & \mathbf{0} \\
                      \mathbf{0} & \mathbf{Z}_2 & \cdots & \mathbf{0} & \mathbf{0} \\
                      \vdots & \vdots & \ddots\: & \vdots & \vdots \\
                      \mathbf{0} & \mathbf{0} & \cdots & \mathbf{Z}_Q & \mathbf{0} \\
                      \mathbf{0} & \mathbf{0} & \cdots & \mathbf{0} & \mathbf{Z}_{Q+1}
                    \end{matrix}
 \right],  
 \end{equation*}
where, for $i \in {[Q]}$, the matrix $\mathbf{Z}_i$, of dimension $\lambda_i \times T$, is constructed as follows
\begin{equation}
 \nonumber
 \mathbf{Z}_i = \left[ \begin{matrix} \mathbf{0}_{\lambda_i \times (i-1) \left\lceil \frac{T}{k} \right\rceil} & \bar{\mathbf{Z}}_i &  \mathbf{0}_{\lambda_i \times (Q-i) \left\lceil \frac{T}{k} \right\rceil} & \mathbf{0}_{\lambda_i \times T_{\text{rem}}} \end{matrix} \right],
\end{equation}
where $\bar{\mathbf{Z}}_i$, of dimension $\lambda_i \times \left\lceil \frac{T}{k} \right\rceil$, has as rows all non-zero vectors of length $\left\lceil \frac{T}{k} \right\rceil$. Therefore, $\lambda_i = 2^{\lceil T/k \rceil}-1$. 
Similarly, the matrix $\mathbf{Z}_{Q+1}$, of dimension $\lambda_{Q+1} \times T$, is constructed as follows
\begin{equation}
 \nonumber
 {\mathbf{Z}_{Q+1}} = \left[ \begin{matrix} \mathbf{0}_{\lambda_{Q+1} \times Q \left\lceil \frac{T}{k} \right\rceil} & {\bar{\mathbf{Z}}_{Q+1}} \end{matrix} \right],
\end{equation}
where {$\bar{\mathbf{Z}}_{Q+1}$,} of dimension ${\lambda_{Q+1}} \times T_{\text{rem}}$, has as rows all non-zero vectors of {length $T_{\text{rem}}$}. Therefore, $\lambda_{Q+1} = 2^{T_{\text{rem}}}-1$.

In other words, the matrix $\mathbf{P}$ is constructed as a block-diagonal matrix, with the diagonal elements being {$\bar{\mathbf{Z}}_i$} for all $i \in {[Q+1]}$. Therefore, equation \eqref{theorem_ub1} holds by computing 
\begin{equation}
\nonumber
 T_k = \sum\limits_{i=1}^{Q+1} \lambda_i = {Q}\left(2^{\left\lceil \frac{T}{k} \right\rceil} -1 \right) + 2^{T_{\text{rem}}} - 1 \leq k2^{\left\lceil \frac{T}{k} \right\rceil}.
\end{equation}

What remains is to show that any vector {$\mathbf{g}_i, i \in [n],$} 
can be reconstructed by adding at most $k$ vectors of $\mathbf{P}$. 
To show this, {we prove that any vector $\mathbf{v} \in \mathbb{F}_2^T$  can indeed be constructed with the proposed design of $\mathbf{P}$. We note that} we can express {the vector $\mathbf{v}$} as $\mathbf{v} = \left[\mathbf{v}_1 \: \cdots \: \mathbf{v}_{Q+1} \right]$, where $\mathbf{v}_i, i \in {[Q]}$ are parts of the vector $\mathbf{v}$ each of length $ \left\lceil \frac{T}{k} \right\rceil $, while $\mathbf{v}_{Q+1}$ is the last part of $\mathbf{v}$ of length $T_{\text{rem}}$.  Then, we can write $ \mathbf{v} = \sum\limits_{i \in \mathcal{K}(\mathbf{v})} \bar{\mathbf{v}}_i,$
where $\bar{\mathbf{v}}_i = \left[ \mathbf{0}_{(i-1) \left\lceil \frac{T}{k} \right\rceil} \quad \mathbf{v}_i \quad \mathbf{0}_{(Q-i) \left\lceil \frac{T}{k} \right\rceil} \: \mathbf{0}_{T_{\text{rem}}} \right]$ for $i \in {[Q]}$, $\bar{\mathbf{v}}_{Q+1} = \left[ \mathbf{0}_{{Q} \left\lceil \frac{T}{k} \right\rceil} \quad \mathbf{v}_{Q+1} \right]$ and $\mathcal{K}(\mathbf{v}) \subseteq {[Q+1]}$ is the set of indices for which $\mathbf{v}_i$ is not all-zero. According to the construction of $\mathbf{P}$, for all $i \in \mathcal{K}(\mathbf{v})$, the corresponding vector $\mathbf{v}_i$ is one of the rows in {$\mathbf{Z}_i$.} The proof concludes by noting that $|\Set{K}(\mathbf{v})| \leq k $.
This is true because, if $k$ does not divide $T$, then $Q \leq k-1$; otherwise, $Q = k$ but $T_{\text{rem}} = 0$ ({\it i.e.,} $\mathbf{v}_{Q+1}$ does not exist), therefore ${\Set{K}(\mathbf{v})} \subseteq [k]$.
This completes the proof of the upper bound in~\eqref{theorem_ub1}.

 \textbf{Example:} We show how the scheme works via a small example, where $T = 8$ and $k = 3$. 
For this particular example, we have $Q = \left\lfloor T/ \left\lceil \frac{T}{k} \right\rceil \right\rfloor = 2$ and $T_{\text{rem}} = T - Q\left\lceil \frac{T}{k}\right\rceil = 2$. Thus,
the idea is that, to reconstruct a vector {$\mathbf{v}\in \mathbb{F}_2^8$}, we treat {$\mathbf{v}$} as $k=3$ disjoint parts; the first $2$ are of length $\left\lceil \frac{T}{k} \right\rceil = 3$ and the remaining part is of length
{$T_{\text{rem}} =2$.}
We then construct $\mathbf{P}$ as $k = 3$ disjoint sections, where each section allows us to reconstruct one part of the vector. Specifically, we {construct}
 \begin{align*}
 {\bar{\mathbf{Z}}_1 = \bar{\mathbf{Z}}_2} = \left[\begin{matrix}
                                                  0 & 0 & 1 \\
                                                  0 & 1 & 0 \\
                                                  0 & 1 & 1 \\
                                                  1 & 0 & 0 \\
                                                  1 & 0 & 1 \\
                                                  1 & 1 & 0 \\
                                                  1 & 1 & 1
                                                 \end{matrix}
 \right], \
 &{\bar{\mathbf{Z}}_3} = \left[\begin{matrix}
                                                  0 & 1 \\
                                                  1 & 0 \\
                                                  1 & 1
                                                 \end{matrix}
 \right],
 &{\mathbf{P}} = \left[ \begin{matrix}
                       {\bar{\mathbf{Z}}_1} & \mathbf{0}_{7 \times 3} & \mathbf{0}_{7 \times 2} \\
                       \mathbf{0}_{7 \times 3} & {\bar{\mathbf{Z}}_2} & \mathbf{0}_{7 \times 2} \\
                       \mathbf{0}_{3 \times 3} &  \mathbf{0}_{3 \times 3} & {\bar{\mathbf{Z}}_3}
                      \end{matrix}
 \right].
 \end{align*}
 Any vector $\mathbf{v}$ can be reconstructed by picking at most $k$ vectors out of $\mathbf{P}$, one from each section. For example, let $\mathbf{v} = \left[ 0 \: 1 \: 0 \: 0 \: 1 \: 1 \: 1 \: 0 \right]$. This vector can be reconstructed by adding vectors number $2$, $10$ and $16$ from $\mathbf{P}$.

\section{Constructions for small values of $n$ and $k$}
\label{sec::BipartiteGraphRep}

\begin{figure*}
 \centering
 \subfigure[$k = 2$]{
 \includegraphics[width=0.45\columnwidth]{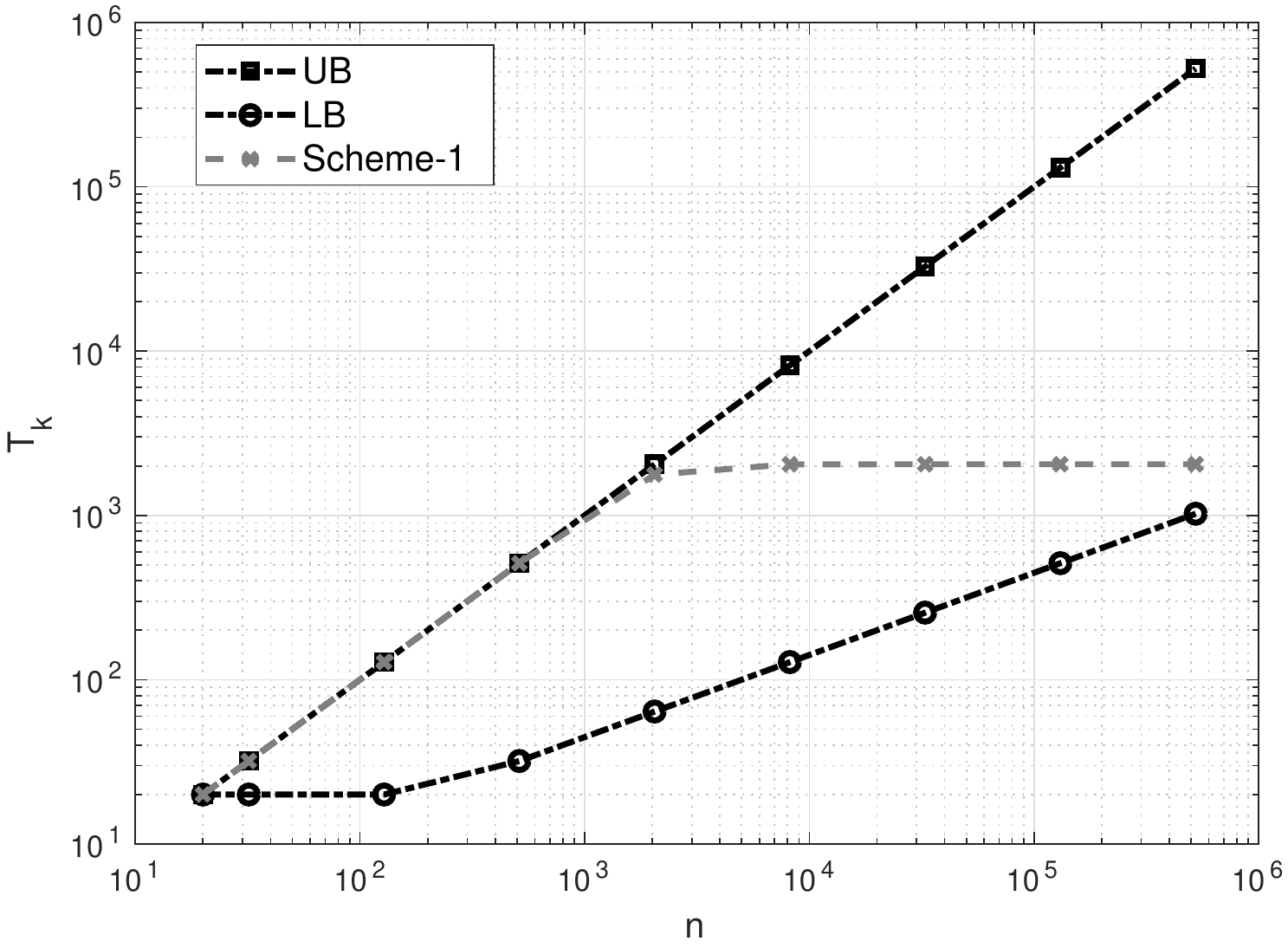}
 \label{fig::Comp2}
 }
 \subfigure[$k=5$]{
 \includegraphics[width=0.45\columnwidth]{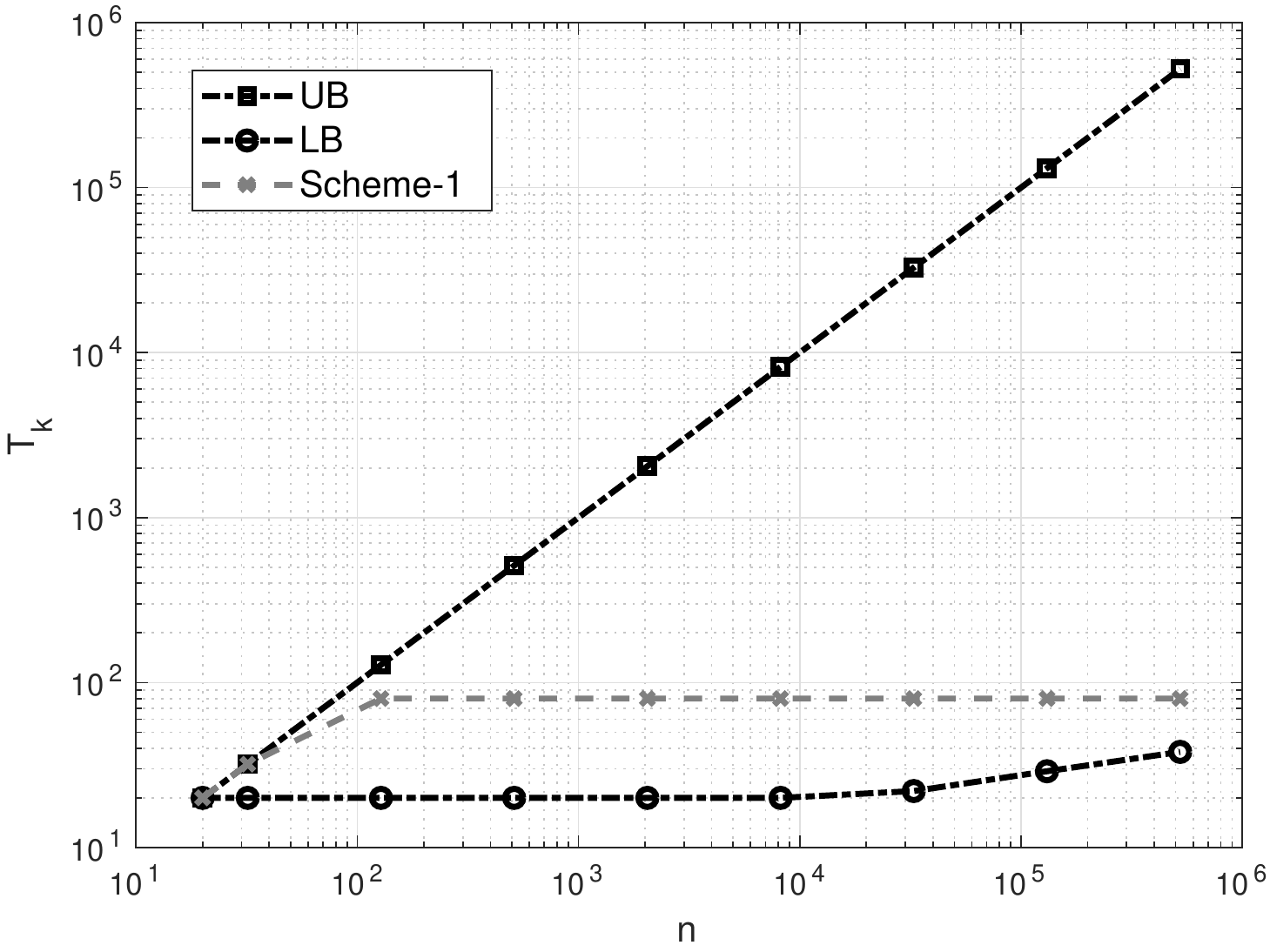}
 \label{fig::Comp5}
 }
 \caption{Performance of the scheme in Theorem~\ref{theorem_main} (referred to as Scheme-1) for different values of $n$, compared against the lower bound LB in equation~\eqref{eq::lb} and the upper bound UB of sending uncoded transmissions - $T = 20$.}
 \label{fig::Comp}
 \end{figure*}
In Section~\ref{sec:MainRes}, we have proved that, independently of the value of $n$, if $k \geq \lceil T/2 \rceil$, then it is sufficient to add one additional transmission to the $T$ transmissions of the conventional index coding scheme. Moreover,
the analysis provided in Lemma~\ref{lemma::SpecCase} showed the order-optimality of our universal scheme in Theorem~\ref{theorem_main} (referred to as Scheme-1) for values of $k < \lceil T/2 \rceil$ when $n$ is large ({\it i.e.,} exponential in $T$). 
Figure~\ref{fig::Comp} shows the performance of Scheme-1 in Theorem~\ref{theorem_main} as a function of the values of $n$ for $T=20$, with $k = 2$ in Figure~\ref{fig::Comp2} and $k = 5$ in Figure~\ref{fig::Comp5}. The performance of Scheme-1 was obtained by averaging over 1000 random index coding instances. In each instance, a code is constructed using the scheme described in Section~\ref{eq:ConstrUB}, and only the rows actually used by the clients $c_{[n]}$ are retained. The performance of the scheme is finally computed by the average number of rows retained in those 1000 iterations. Figure~\ref{fig::Comp} shows that our proposed scheme performs well not only for the case of large $n$ ({\it i.e.}, $n=2^T-1$) but also for lower values of $n$. However, Figure~\ref{fig::Comp} also suggests that for small values of both $n$ and $k$ (note the left-half of the plot in Figure~\ref{fig::Comp2}),
we need to devise schemes that better adapt to the specific values of the index coding matrix $\mathbf{A}$ and vectors {$\mathbf{g}_i, i \in [n]$ (recall that Scheme-1 is universal, and hence independent of the value of $\mathbf{A}$).} We next propose and analyze the performance of such algorithms.

\subsection{Special Instances}
\label{subsec:SpecInst}
We first represent the problem through a bipartite graph as follows. We assume that the rank of the matrix $\mathbf{G}$ is $T$. Then, there exists a set of $T$ linearly independent vectors in $\mathbf{G}$; without loss of generality, we denote them as $\mathbf{g}_1$ to $\mathbf{g}_T$. Therefore, each vector $\mathbf{g}_{i+T}, i \in [n-T],$ can be expressed as a linear combination of some/all vectors from $\mathbf{g}_{[T]}$; we denote these vectors as the component vectors of $\mathbf{g}_{i+T}$.
We can then represent the problem as a bipartite graph $(\Set{U} \cup \Set{V}, \Set{E})$ with $| \Set{U}| = T$ and $| \Set{V}| = n-T$, where $u_i \in \Set{U}$ represents the vector $\mathbf{g}_{i}$ for $i \in [T]$, ${v_j} \in \Set{V}$ represents the vector {$\mathbf{g}_{j+T}$ for $j \in [n-T]$,} and an edge exists from node $u_i$ to node $v_j$ if $\mathbf{g}_i$ is one of the component vectors of $\mathbf{g}_{j+T}$. Figure~\ref{fig::bipartitegraphrep} shows an example of such graph, where $n = 9$ and $T = 6$. 
For instance, $v_1$ ({\it i.e.,} $\mathbf{g}_7$) can be reconstructed by adding $u_i, i \in [4]$ ({\it i.e.,} $\mathbf{g}_i, i \in [4]$).
Given a node $s$ in the graph, we refer to the sets $\Set{O}_s$ and $\Set{I}_s$ as the {\it outbound} and {\it inbound} sets of $s$, respectively: the inbound set contains the nodes which have edges outgoing to node $s$, and the outbound set contains the nodes to which node $s$ has outgoing edges ({\it i.e.}, the nodes each of which has an incoming edge from $s$). Nodes on either sides of the bipartite graph have either inbound or outbound sets.
For instance, with reference to Figure~\ref{fig::bipartitegraphrep}, $\Set{O}_{u_1} = \{v_1,v_2,v_3 \}$ and $\Set{I}_{v_1} = \{u_1,u_2,u_3,u_4 \}$.
For this particular example, there exists a scheme with $T_2 = 6$ which can reconstruct any vector with at most $k=2$ additions. The matrix $\mathbf{A}_2$ which corresponds to this solution consists of the following vectors: $\mathbf{g}_1$, $\mathbf{g}_1 + \mathbf{g}_2$, $\mathbf{g}_1 + \mathbf{g}_2 + \mathbf{g}_3$, $\mathbf{g}_1 + \mathbf{g}_2 + \mathbf{g}_3 + \mathbf{g}_4$, $\mathbf{g}_5$ and $\mathbf{g}_5 + \mathbf{g}_6$.
It is not hard to see that each vector in $\mathbf{G}$ can be reconstructed by adding at most $2$ vectors in $\mathbf{A}_2$. The vectors in $\mathbf{A}_2$ that are not in $\mathbf{G}$ can be aptly represented as intermediate nodes on the previously described bipartite graph. These intermediate nodes are shown in Figure \ref{fig::bipartitegraphrepsol} as highlighted nodes. Each added node represents a new vector, which is the sum of the vectors associated to the nodes in its inbound set. We refer to the process of adding these intermediate nodes as creating a {\it branch}, which is defined next.

\begin{definition} Given an ordered set $\Set{S} = \{s_1, \: \cdots, \: s_S \}$ of nodes,
where $s_i$ precedes $s_{i+1}$ for $i \in [S-1]$, 
a {\it branch on $\Set{S}$} is a set $\Set{S}^\prime = \{ s^\prime_1, \: \cdots, \: s^\prime_{S-1} \}$ of $S-1$ intermediate nodes added to the graph with the following connections: node $s^\prime_1$ has two incoming edges from $s_1$ and $s_2$, and for $i \in [S-1] \setminus \{1\}$, $s^\prime_i$ has two incoming edges from nodes $s^\prime_{i-1}$ and $s_{i+1}$.
\end{definition}

For the example in Figure \ref{fig::bipartitegraphrepsol}, we created branches on two ordered sets, $\Set{S}_1 = \{u_1, \: u_2, \: u_3, \: u_4 \}$ and $\Set{S}_2 = \{u_5, \: u_6 \}$. Once the branch is added, we can change the {connections} of the nodes in $\Set{V}$ in accordance to the added vectors. For the example in Figure \ref{fig::bipartitegraphrepsol}, we can replace $u_{[4]}$ in $\Set{I}_{v_1}$ with only $s_{3}$.
\begin{figure}
  \centering
  \begin{minipage}{0.4\columnwidth}
  \centering
  \includegraphics[width=0.9\textwidth]{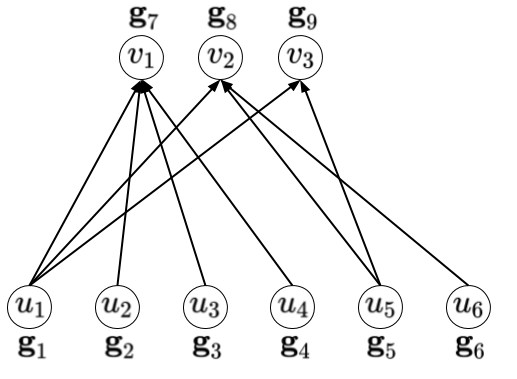}
  \caption{Bipartite graph representation.}
  \label{fig::bipartitegraphrep}
    \end{minipage}
      \begin{minipage}{0.4\columnwidth}
  \centering
  \includegraphics[width=0.9\textwidth]{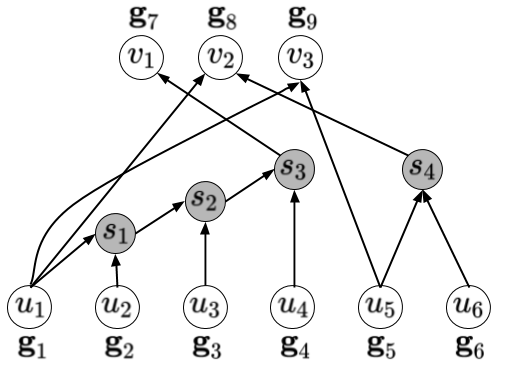}
  \caption{Optimal representation when $k=2$.}
  \label{fig::bipartitegraphrepsol}
    \end{minipage}
\end{figure}
Using this representation, we have the following lemma.


\begin{lemma}
\label{lem::subsetlem}
 If $\mathcal{O}_{u_{i_T}} \subseteq \mathcal{O}_{u_{i_{T-1}}} \subseteq \cdots \subseteq \mathcal{O}_{u_{i_1}}$ for some permutation $i_1, \cdots, i_T$ of $[T]$, then this instance can be solved by exactly $T$ transmissions for any $k \geq 2$.
\end{lemma}

\begin{Pf}
 One solution of such instance would involve creating a branch on the set $\Set{S} = \{u_{i_1}, \: u_{i_{2}}, \: \cdots, \: u_{i_T} \}$.
 The scheme used would have the matrix $\mathbf{A}_2$ with its $t$-th row $\mathbf{a}_t = \sum\limits_{\ell = 1}^{t} \mathbf{g}_{i_\ell}$ for $t \in [T]$. Note that $\mathbf{g}_{i_1} = \mathbf{a}_{1}$ and $\mathbf{a}_{t} + \mathbf{a}_{t-1} = \mathbf{g}_{i_t}$ for all $t \in [T]\setminus\{1\}$. 
 Moreover, for $j \in [n]\setminus [T]$, if $v_{j-T} \in \mathcal{O}_{u_{i_t}}$ for some $i_t$, then $ v_{j-T} \in \mathcal{O}_{u_{i_\ell}}$ for all $\ell \leq t$. 
If we let $t$ be the maximum index for which $v_{j-T} \in \mathcal{O}_{u_{i_t}}$, then we have $\Set{I}_{v_{j-T}} = \{u_{i_1}, \: \cdots, \: u_{i_t}\}$, and so we get $\mathbf{g}_{j} = \sum\limits_{\ell = 1}^t \mathbf{g}_{i_\ell} = \mathbf{a}_t$. This completes the proof.
\end{Pf}

\begin{corollary}
\label{cor::circuit}
 For $\mathbf{G} \in \mathbb{F}_2^{n \times T}$ of rank $T$, if $n = T+1$, then this instance can be solved in $T$ transmissions for any $k \geq 2$.
\end{corollary}
\begin{Pf} Without loss of generality, let $\mathbf{g}_{[T]}$ be a set of linearly independent vectors of $\mathbf{G}$. 
Then, we have $\Set{O}_{u_i} = \{v_1\}$ for $i \in \Set{I}_{v_1}$ and $\Set{O}_{u_j} = \emptyset$ for $j \in [T]\setminus \Set{I}_{v_1}$. Thus, from Lemma~\ref{lem::subsetlem}, this instance can be solved in $T$ transmissions.
This completes the proof.
\end{Pf}

\subsection{Algorithms for General Instances}
\label{sec::algorithms}

We here propose two different algorithms, namely Successive Circuit Removing (SCR) and Branch-Search, and analyze their performance.
  
\noindent\textbf{Algorithm 1: Successive Circuit Removing (SCR).} 
Our first proposed algorithm is based on Corollary~\ref{cor::circuit}, which can be interpreted as follows: any matrix $\mathbf{G}$ of $r+1$ row vectors and rank $r$ can be reconstructed by a corresponding $\mathbf{A}_2$ matrix with $r$ rows. 
{If there does not exist any subset of rows of $\mathbf{G}$ with rank less than $r$, we call $\mathbf{G}$ {\it a circuit}}\footnote{This is in accordance to the definition of a circuit for a matroid\cite{oxley2006matroid}.}.
Our algorithm works for the case $k = 2^q$, for some integer $q$. We first describe SCR for the case where $q=1$, and then extend it to general values of $q$.
{The algorithm works as follows:}

\medskip

\noindent $1)$ \textit{Circuit Finding:} find a set of vectors of $\mathbf{G}$ that form a circuit of small size. Denote the size of this circuit as $r+1$.\\
\noindent $2)$ \textit{Matrix Update:} apply Corollary \ref{cor::circuit} to find a set of $r$ vectors that can optimally reconstruct the circuit by adding at most $k=2$ of them, and add this set to $\mathbf{A}_2$.\\
\noindent $3)$ \textit{Circuit Removing:} update $\mathbf{G}$ by removing the circuit. Repeat the first two steps until the matrix $\mathbf{G}$ is of size $T^\prime \times T$ and of rank $T^\prime$, where $T^\prime \leq T$. Then, add these vectors to $\mathbf{A}_2$.

\medskip

Once SCR is executed, the output is a matrix $\mathbf{A}_2$ such that any vector in $\mathbf{G}$ can be reconstructed by adding at most $k=2$ vectors of $\mathbf{A}_2$. Consider now the case where $q=2$ ({\it i.e.,} $k=4$) for example. In this case, a second application of SCR on the matrix $\mathbf{A}_2$ would yield another matrix, denoted as $\mathbf{A}_4$, such that any row in $\mathbf{A}_2$ can be reconstructed by adding at most $2$ vectors of $\mathbf{A}_4$. Therefore, any vector in $\mathbf{G}$ can now be reconstructed by adding at most $4$ vectors of $\mathbf{A}_4$. We can therefore extrapolate this idea for a general $q$ by successively applying SCR $q$ times on $\mathbf{G}$ to obtain $\mathbf{A}_k$, with $k=2^q$.

The following theorem gives a closed form characterization of the best and worst case performance of SCR.

\begin{theorem}
 \label{thm::SCR}
 Let $T_q^{\text{SCR}}$ be the number of vectors in $\mathbf{A}_k$ obtained via SCR. Then, for $k=2^q$ and integer $q$, we have
 \begin{equation}
  \label{eq::SRCPerf}
  \underbrace{f^{\text{Best}}(f^{\text{Best}}( \cdots f^{\text{Best}}(n)))}_{q \text{ times}} \!\leq\! T_q^{\text{SCR}} \!\leq \! \underbrace{f^{\text{Worst}}(f^{\text{Worst}}( \cdots f^{\text{Worst}}(n)))}_{q \text{ times}},
 \end{equation}
 where $f^{\text{Best}}(n) = 2\left\lfloor \frac{n}{3}\right\rfloor$ and $f^{\text{Worst}}(n) = T \left( \left\lfloor \frac{n}{T+1} \right\rfloor + 1\right)$.
\end{theorem}
\begin{Pf}
First we focus on the case $q = 1$.
The lower bound in~\eqref{eq::SRCPerf} corresponds to the best case when the matrix $\mathbf{G}$ can be partitioned into disjoint circuits of size $3$. In this case, if SCR finds one such circuit in each iteration, then each circuit is replaced with $2$ vectors in $\mathbf{A}_2$ according to Corollary~\ref{cor::circuit}. To obtain the upper bound, note that any collection of $T+1$ has at most $T$ independent vectors, and therefore contains a circuit of at most size $T+1$. Therefore, the upper bound corresponds to the case where the matrix $\mathbf{G}$ can be partitioned into circuits of size $T+1$ and an extra $T$ linearly independent vectors. In that case, the algorithm can go through each of these circuits, adding $T$ vectors to $\mathbf{A}_2$ for each of these circuits, and then add the last $T$ vectors in the last step of the algorithm. Finally, the bounds in \eqref{eq::SRCPerf} for a general $q$ can be proven by a successive repetition of the above arguments.
\end{Pf}

\noindent\textbf{Algorithm~2: Branch-Search.}
A naive approach to determining the optimal matrix $\mathbf{A}_k$ is to consider the whole space $\mathbb{F}_2^T$, loop over all possible subsets of vectors of $\mathbb{F}_2^T$ and, for every subset, check if it can be used as a matrix $\mathbf{A}_k$. The minimum-size subset which can be used as $\mathbf{A}_k$ is indeed the optimal matrix. However, such algorithm requires in the worst case $O\left (2^{2^T} \right )$ number of operations, which makes it prohibitively slow even for very small values of $T$. 
Instead, {the heuristic that we here propose} finds a matrix $\mathbf{A}_k$ more efficiently than the naive search scheme. The main idea behind the heuristic is based on providing a subset $\Set{R} \subset \mathbb{F}_2^T$ which is much smaller than $2^T$ and is guaranteed to have at least one solution. The heuristic then searches for a matrix $\mathbf{A}_k$ by looping over all possible subsets of $\Set{R}$. Our heuristic therefore consists of two sub-algorithms, namely Branch and Search. 
Branch takes as input $\mathbf{G}$, and produces as output a set of vectors $\Set{R}$ which contains at least one solution $\mathbf{A}_k$. The algorithm works as follows:

\medskip

\noindent 1) Find a set of $T$ vectors of $\mathbf{G}$ that are linearly independent. Denote this set as $\Set{B}$.\\
\noindent 2) Create a bipartite graph representation of $\mathbf{G}$ as discussed in Section~\ref{subsec:SpecInst}, using $\Set{B}$ as the independent vectors for $\Set{U}$.\\
\noindent 3) Pick the dependent node $v_i$ with the highest degree, and split ties arbitrarily. Denote by $\text{deg}(v_i)$ the degree of node $v_i$.\\
\noindent 4) Consider the inbound set $\Set{I}_{v_i}$, and sort its elements in a descending order according to their degrees. Without loss of generality, assume that this set of ordered independent nodes is $\Set{I}_{v_i} = \{u_1, \: u_2, \: \cdots , \: u_{\text{deg}(v_i)} \}$.\\
\noindent 5) Create a branch on $\Set{I}_{v_i}$. Denote the new branch nodes as $\{u^\star_1, \: u^\star_2, \: \cdots , \: u^\star_{\text{deg}(v_i)} \}$.\\
\noindent 6) Update the connections of all dependent nodes in accordance with the constructed branch. This is done as follows: for each node $v_j \in \Set{V}$ with $\text{deg}(v_j) \geq k$, if $\Set{I}_{v_j} \cap \Set{I}_{v_i}$ is of the form $\{u_1, \: u_2, \: \cdots, \: u_{\ell}\}$ for some $\ell \leq \text{deg}(v_i)$, then replace $\{u_1, \: u_2, \: \cdots, \: u_{\ell}\}$ in $\Set{I}_{v_j}$ with the single node $u^\star_{\ell}$. Do such replacement for the maximum possible value of $\ell$.\\
\noindent 7) Repeat 3) to 6) until all nodes in $\Set{V}$ have degree at most $k$.

\medskip

The output $\Set{R}$ is the set of vectors corresponding to all nodes in the graph. The next theorem shows that $\Set{R}$ in fact contains one possible $\mathbf{A}_k$, and characterizes the performance of Branch.

\begin{theorem}
\label{thm::branch}
 For a matrix $\mathbf{G}$ of dimension $n \times T$, (a) Branch produces a set $\Set{R}$ which contains at least one possible $\mathbf{A}_k$, (b) the worst-case time complexity $t_{\text{Branch}}$ of Branch is $O(n^2)$, and (c) $|\Set{R}| \leq (n-T)T$.
\end{theorem}
\begin{Pf}
To see (a), note that the algorithm terminates when all dependent nodes have a degree of $k$ or less. In every iteration of the algorithm, the degrees of all dependent nodes either remain the same or are reduced. In addition, at least one dependent node is updated and its degree is reduced to $1$. Therefore the algorithm is guaranteed to terminate. Since all dependent nodes have degrees $k$ or less, their corresponding vectors can be reconstructed by at most $k$ vectors in $\Set{R}$. Therefore, $\Set{R}$ contains at least one solution $\mathbf{A}_k$.

To prove (b), the worst-case runtime of Branch corresponds to going over all nodes in $\Set{V}$, creating a branch for each one. For the $i$-th node considered by Branch, the algorithm would update the dependencies of all dependent nodes with degrees greater than $k$, which are at most $n-i$ nodes. Therefore $ t_{\text{Branch}} = \sum\limits_{i=0}^{n-1} (n-i) = n(n-1) = O(n^2)$.
 
 To prove (c), note that $|\Set{R}|$ is equal to the total number of nodes in all branches created by the algorithm. Therefore we can write $|\Set{R}| \leq \sum\limits_{v_i \in \Set{V}} \text{deg}(v_i) \leq (n-T)T = O(nT)$.
\end{Pf}

Let $t_{\text{Search}}$ be the worst-time complexity of the Search step in Branch-Search. Then the worst-case time complexity of Branch-Search is equal to $t_{\text{BS}} = t_{\text{Branch}} + t_{\text{Search}} \leq O(n^2) + 2^{|\Set{R}|} = O(n^2) + O(2^{nT}) = O(2^{nT})$, which is exponentially better than the complexity of the naive search. Although our heuristic is still of exponential runtime complexity, we observe from numerical simulations that $|\Set{R}|$ is usually much less than $(n-T)T$. Finding more efficient ways of searching through the set $\Set{R}$ to find a solution $\mathbf{A}_k$ is an open question.

\subsection{Numerical Evaluation}
\label{sec::evaluation}

We here explore the performance of our proposed schemes through numerical evaluations. Specifically, we assess the performance in terms of $T_k$ {of 
SCR} and Branch-Search (labeled {as} BS). We compare their performance against the lower bound in equation~\eqref{eq::lb} (labeled as LB), and the upper bound of sending uncoded transmissions ({labeled as} UB).
In particular, we are interested in regimes for which $k < \lceil T/2 \rceil$, because otherwise we know from Theorem~\ref{theorem_main} that $T_k = T+1$. Moreover, we consider values of $n < 2^T-1$, because if $n=2^T-1$ we know from {Lemma~\ref{lemma::SpecCase}} that Scheme-1 is order {optimal. 
For SCR,} we evaluate its average performance (averaged over $1000$ iterations) as well as its upper and lower {bounds} performance established in Theorem~\ref{thm::SCR}. For Branch-Search, we evaluate its average performance (averaged over $1000$ iterations).
Figure~\ref{fig::algperf} shows the performance of all the aforementioned schemes for $T=6$ and $k=2$.
As can be seen {from Figure~\ref{fig::algperf},
SCR consistently} performs better than uncoded transmissions.
In addition, although the current implementation of SCR greedily searches for a small circuit to remove, more sophisticated algorithms for small circuit finding could potentially improve its performance. However, the bounds in~\eqref{eq::SRCPerf} suggest that the performance of SCR is asymptotically $O(n)$. Branch-Search appears to perform better than other schemes in the average sense. Understanding its asymptotic behavior in the worst-case is an interesting open problem.

\begin{figure}
 \centering
   \includegraphics[width=0.65\columnwidth]{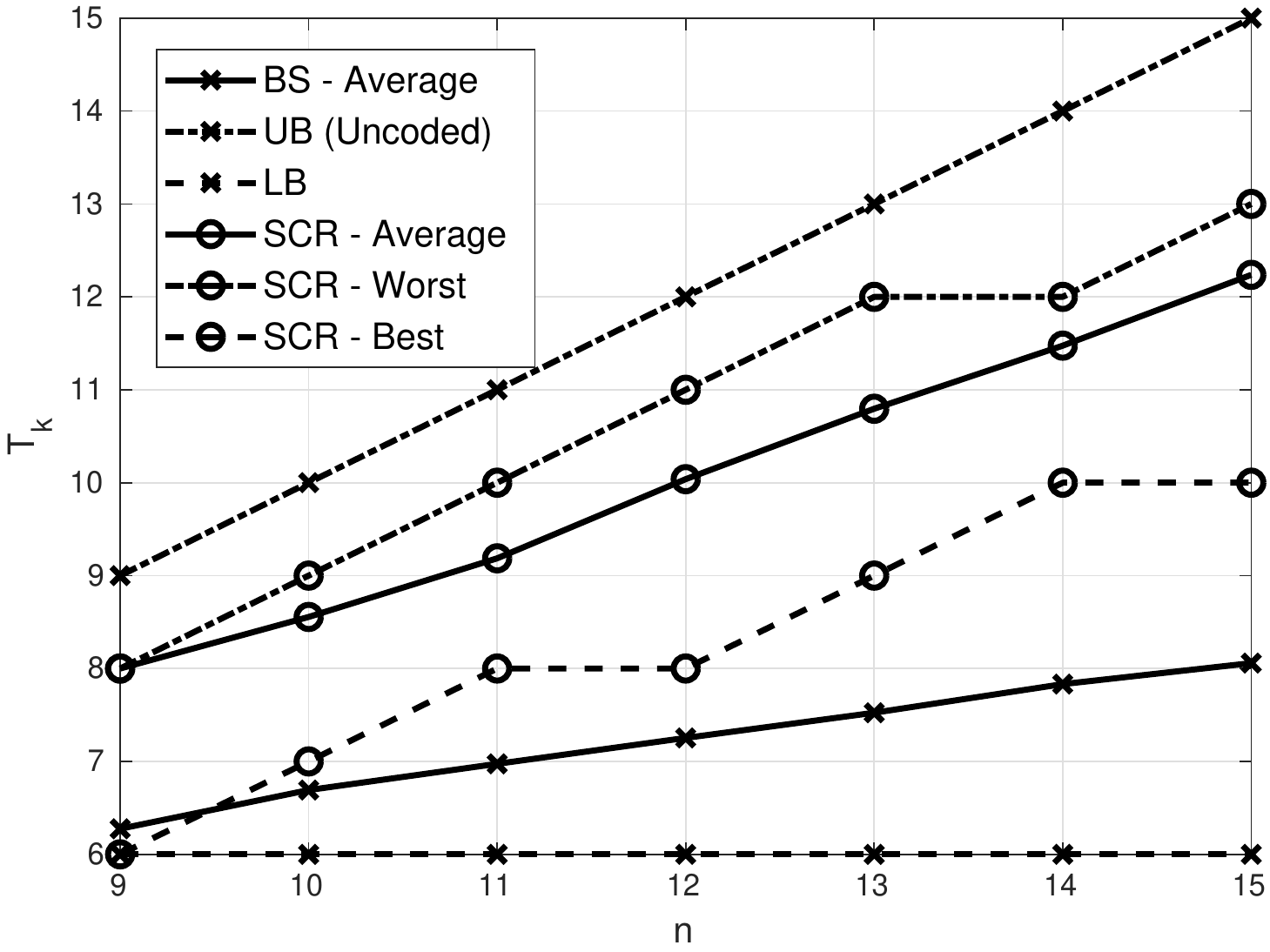}
 \caption{Performance comparison for different schemes - $T = 6$, $k=2$.}
 \label{fig::algperf}
 \end{figure}

\section{Related Work} \label{sec::relatedWork}
Index coding was introduced in~\cite{bar2011index}, where the problem was proven to be NP-hard. 
Given this, several works have aimed at providing
approximate algorithms for the index coding problem~\cite{esfahanizadeh2014matrix,blasiak2010index,chaudhry2008efficient}. 
{In our work, we were interested in studying} the index coding problem from the perspective of private information delivery.

The problem of protecting privacy was initially proposed
to enable the disclosure of databases for public access, while maintaining the anonymity of the clients~\cite{aggarwal2008general}.
Similar concerns have been raised in the context of
{\it Private Information Retrieval} (PIR),
{which was introduced in~\cite{chor1998private} and has received a fair amount of attention~\cite{freij2016private,chen2017capacity,sun2018private,BBB,banawan2}.}
In particular, in PIR the goal is to ensure that no information about {the identity of} clients' requests is revealed to a set of malicious databases when clients are trying to retrieve information from them.
Similarly, the problem of {\it Oblivious Transfer} was studied~\cite{Brassard1987,mishra2014oblivious} to establish, by means of cryptographic techniques, two-way private connections between the clients and the server.
We note that it is not clear how the use of cryptographic approaches would help in our setup.
A curious client, in fact, obtains information about other clients once she learns the transmitted combinations of {the} messages, {\it i.e.,} the coding operations. In other words, given that a curious client has also requested data, she needs to learn how the transmitted messages are coded, in order to be able to decode her own requested message.

We were here interested in addressing privacy concerns in broadcast domains.
In particular, we analyzed this problem within the index coding framework, as we recently proposed in~\cite{karmoose2017private}.
This problem differs from secure index coding~\cite{dau2012security,narayanan2018private}, where the goal is to guarantee that an external eavesdropper (with her own side information set) in~\cite{dau2012security}, and each client in~\cite{narayanan2018private}, does not learn any information about the {\it content} of the messages other than her {requested message.} 
Differently, our goal was to limit the information that a client can learn about the {\it identities} of the requests of other clients (however, the two approaches could be combined).
{1Note that the techniques developed here can fundamentally differ from those designed for secure index coding.
As an extreme example, in fact, the server in our setup can trivially send all the messages that it possesses in an uncoded manner on the broadcast channel. In this case, a curious client will be able to decode all messages, but would still not be able to infer which messages were requested/possessed by other clients, and would learn nothing about their side information. 
This property is what fundamentally contrasts the problem under consideration from the works in~\cite{dau2012security,narayanan2018private}.}
Moreover, our approach here has a significant difference with respect to~\cite{karmoose2017private}.
In fact, while in~\cite{karmoose2017private} our goal was to design the coding matrix to guarantee a high-level of privacy, here we assumed that an index coding matrix (that satisfies all clients) was given to us and we developed methods to increase its achieved level of privacy.

The use of $k$-limited-access schemes allows the server to transform an existing index code into a {\it locally decodable index code}~\cite{haviv2011linear,natarajan2018locally}. 
Locally decodable index codes allow each client to decode her request using at most $k$ symbols out of the codeword, where $k$ is referred to as the locality of the code. 
In~\cite{haviv2011linear}, the authors showed that the optimal scalar linear locally decodable index codes {\karmoose with locality $1$} are the ones obtained from the coloring of the information graph of the index coding problem. In addition, they provided probabilistic {\karmoose results} on the existence (and the impossibility of existence) of locally decodable codes with particular lengths and localities for {\karmoose index coding problems on random graphs}. 
In~\cite{natarajan2018locally}, the authors {\karmoose extended one result in~\cite{haviv2011linear} where} they showed that the optimal {\karmoose \it vector} linear locally decodable index codes with locality $1$ are obtained from the fractional coloring of the information graph.
In addition, they provided a scheme which allows the construction of locally decodable codes for a particular set of index coding instances with special properties, {\it i.e.}, when certain covering properties are maintained on the side information graph of the index coding problem.
Differently from these works, one of the main results of this paper consisted of providing deterministic constructions/schemes which transform any existing index code into an equivalent code with locality $k$. In addition, our schemes are universal, {\it i.e.}, they do not depend on the underlying index coding instance.

The solution that we here proposed to limit the privacy leakage  is based on finding overcomplete bases.
This approach 
is closely related to
compressed sensing and dictionary learning~\cite{chen2016compressed}, where the goal is to learn a dictionary of signals such that other signals can be {\it sparsely} and {\it accurately} represented using atoms from this dictionary.
These problems seek lossy solutions, {\it i.e.,} signal reconstruction is not necessarily perfect.
This allows a convex optimization formulation of the problem, which can be solved efficiently~\cite{rubinstein2010dictionaries}.
{In contrast,} our problem was concerned with lossless reconstructions, in which case the optimization problem is no longer convex.

\section{Conclusion}
\label{sec::conclusion}
In this paper, we studied privacy risks in index coding. {This problem is} motivated by the observation that, {since} the coding matrix needs to be available to all clients, {then} some clients may be able to infer the identity {of the request} and side information of other clients. 
We proposed the use of $k$-limited-access schemes: these schemes transform the coding matrix so that we can restrict each client to {access} at most $k$-rows of the transformed matrix as opposed to the whole of it.
We explored two privacy metrics, one based on entropy arguments, and the other on the maximal information leakage. 
Both metrics indicate that the amount of privacy increases with the number of rows that we hide. We then designed polynomial time {universal $k$-limited-access schemes,} that do not depend on the structure of the index coding matrix $\mathbf{A}$ and proved that they are order-optimal when either $k$ or $n$ is large. For the case where both $k$ and $n$ are small, we proposed algorithms that depend on the structure of the {index coding} matrix  $\mathbf{A}$ and provide improved performance.
We overall found that there exists an {inherent} trade-off between privacy and bandwidth (number of broadcast transmissions),
and that in some cases we can achieve significant privacy with minimal overhead.

\appendices
\section{Proof of Lemma~\ref{lem::subspaceCount}}
\label{app::subsetlen}
 The proof is based on simple counting arguments. A subspace $L$ contains all vectors in $L_n$, the number of which is $2^k$. 
A subspace $L$ therefore consists of a set of $T-k$ linearly independent vectors $\{ v_1, \cdots v_{T-k} \}$ that are in $\mathbb{F}_2^m \setminus L_n$, and all linear combinations of $\{ v_{[T-k]} \}$ and vectors in $L_n$. 
We now enumerate the number of ways such a subspace $L$, with $L_n \subseteq L$, can be constructed.
We first pick a vector $v_1 \in \mathbb{F}_2^m \setminus L_n$. The total number of possible choices for $v_1$ is equal to $2^m - 2^k$. 
Once $v_1$ is selected to be in $L$, then all vectors in $v_1 + L_n$ are added to $L$, where $v_1 + L_n$ is the set of vectors obtained by adding $v_1$ to all possible vectors in $L_n$. Therefore, by picking $v_1$, the total number of vectors of $\mathbb{F}_2^m$ that do not belong to $L$ is now equal to $2^m - 2^{k+1}$, out of which we pick $v_2$. The above process is repeated until all vectors $\{v_{[T-k]} \}$ are selected. Therefore, the total number of such choices becomes $\prod_{\ell=0}^{T-k-1} \left( 2^m-2^{k+\ell} \right)$. In order to compute the total number of subspaces, we need to divide this number by the total number of basis vectors ({\it i.e.,} linearly independent vectors) used to represent the vectors in $L \setminus L_n$; we denote them by $\{b_1, \cdots, b_{T-k} \}$. The number of vectors in such a basis is $T-k$. Given a subspace $L$, we pick $b_1$ from the set of vectors in $L \setminus L_n$, the number of which is $2^T - 2^k$. Then we pick $b_2$ from the set of vectors $L \setminus \left( L_n + b_1 \right) $, the number of which is $2^T - 2^{k+1}$. We repeat the previous argument for all $T-k$ vectors. The total number of such basis vectors is therefore equal to $\prod_{\ell=0}^{T-k-1} \left( 2^T-2^{k+\ell} \right)$. Dividing the two quantities therefore proves Lemma~\ref{lem::subspaceCount}.
 
 \section{Proof of Theorem~\ref{thm::P_MIL}}
 \label{app::P_MIL}
 
 To prove Theorem~\ref{thm::P_MIL}, we first recall the definition of $\mathcal{G}(q_i,\mathcal{S}_i)$. Given $q_i$ and $\Set{S}_i$, $\mathcal{G}(q_i,\mathcal{S}_i) $ is the set which contains all possible $i$-th vectors $\mathbf{g}_i$ of the realization $G$ of the matrix $\mathbf{G}$, namely
\begin{equation}
 \mathcal{G}(q_i,\mathcal{S}_i) = \left\{ \mathbf{g} \in \mathbb{F}_2^{m} \: | \:  g_{q_i} = 1, g_{[m] \setminus \{ q_i \cup \mathcal{S}_i \}} = 0 \right\}. \nonumber
\end{equation}
In addition, we define the following set. Given $\mathbf{g}_i$ and an integer $r$, we let $\mathcal{D}(\mathbf{g}_i,r)$ be the set of all possible matrices $\mathbf{A}_k^{(i)}$ of $r$ rows from which $\mathbf{g}_i$ can be reconstructed, namely
\begin{equation*}
 \mathcal{D}(\mathbf{g}_i,r) = \left\{ \mathbf{Z} \in \mathbb{F}_2^{r \times m} \: | \: \exists \mathbf{d} \in \mathbb{F}_2^r \text{ s.t. }  \mathbf{g}_i = \mathbf{d} \mathbf{Z} \right\}.
\end{equation*}
Note that the definition of $\mathcal{D}(\mathbf{g}_i,r)$ is different than that of $\mathcal{P}(\mathbf{g}_i,\mathbf{A}_k,r)$ in that it is not dependent on a specific matrix $\mathbf{A}_k$.
Then, we can write

\begin{align*}
 P_k^{(\text{MIL})} =\mathcal{L}(A \rightarrow A_k^{(n)} | Q_n=q_n, S_n = \mathcal{S}_n) &\stackrel{{\rm{(a)}}}{\leq} \log \left| A_k^{(n)} | Q_n=q_n, S_n = \mathcal{S}_n \right| \\
 &\stackrel{{\rm{(b)}}}{=} \log \left| \bigcup\limits_{r=1}^k \bigcup\limits_{\mathbf{g}_n \in \mathcal{G}(q_n,\mathcal{S}_n)} \mathcal{D}(\mathbf{g}_n,r) \right| \\
 &\leq \log \left ( \sum\limits_{r=1}^{k} \sum\limits_{\mathbf{g}_n \in \mathcal{G}(q_n,\mathcal{S}_n)} \left|\mathcal{D}(\mathbf{g}_n,r) \right| \right )\\
 &\stackrel{{\rm{(c)}}}{=} \log \left( 2^{|\Set{S}_n|} \sum\limits_{r=1}^k |\mathcal{D}(\mathbf{g}^\prime_n,r)| \right) \\
 &\stackrel{{\rm{(d)}}}{\leq} \log \left(2^{|\Set{S}_n|} \sum\limits_{r=1}^{k} \prod\limits_{j=0}^{r-2} (2^m - 2^{j+1} )\right ) \\
 &\leq \log \left( 2^{|\Set{S}_n|} k (2^m - 2)^{k-1} \right ) 
\\&= O(|\Set{S}_n| + mk) ,
\end{align*}
where: {(i) the equality in} ${\rm{(a)}}$ follows from {Property~2} of the MIL;
{\karmoose (ii)} the equality in ${\rm{(b)}}$ follows by noting that, given $Q_n$ and $S_n$, a possible $A_k^{(n)}$ would belong to $\Set{D}(\mathbf{g}_n,r)$ for some $r \in [k]$ and some $\mathbf{g}_n \in \Set{G}(Q_n, S_n)$;
{\karmoose (iii)} the equality in ${\rm{(c)}}$ follows by noting that, by symmetry, the number of matrices with $r$ rows from which the vector $\mathbf{g}_i$ can be reconstructed is the same for every possible vector $\mathbf{g}_i \in \Set{G}(q_i,\Set{S}_i)$. Therefore, the sum over {$\mathbf{g}_n$} can be replaced by {$\Set{D}(\mathbf{g}_n^\prime,r) \times |\mathcal{G}(q_n,\mathcal{S}_n)|$} where $\mathbf{g}^{\prime}_n$ is any arbitrary vector in {$\mathcal{G}(q_n,\mathcal{S}_n)$.} Based on the structure of the vectors $\karmoose \mathbf{g}_n \in \Set{G}(q_n,\Set{S}_n)$, {\it i.e.}, one in position $\karmoose q_n$ and zeros in the positions $[m]\setminus \{{\karmoose q_n \cup \Set{S}_n}\}$, it {follows} that $|\Set{G}({\karmoose q_n,\Set{S}_n})| = 2^{|{\karmoose \Set{S}_n}|}$;
(iv) the inequality in
${\rm{(d)}}$ is obtained by counting arguments similar to {those in the proof of} {Lemma~\ref{lem::subspaceCount}. In particular, we enumerate the number of ways we can construct a matrix $\mathbf{A}_k^{(n)}$ with $r$ linearly independent rows, {which {\karmoose when} linearly combined} gives $\mathbf{g}_i$. We first pick a row vector $v_1 \in \mathbb{F}_2^m \setminus \text{Span}(\mathbf{g}_i)$, where $\text{Span}(\Set{X})$ of a set of row vectors $\Set{X}$ is the row span of these vectors; the number of possible vectors $v_1$ is $2^m - 2$. Then, we pick a second row vector $v_2 \in \mathbb{F}_2^m \setminus \text{Span}(\{ \mathbf{g}_i, v_1\})$; the number of possible vectors $v_2$ is $2^m - 2^2$. We repeat this argument for $r-1$ vectors; the $r$-th vector is then selected so that {a linear combination} of all $r$ vectors is {equal to $\mathbf{g}_i$.}

\section{Proof of Theorem~\ref{thm::P_MIL_Conv}}
\label{app::P_MIL_Conv}
We have
 \begin{align*}
 \bar{P}_k^{(\text{MIL})} = \mathcal{L}(A \rightarrow A | {Q_n=q_n,S_n=\mathcal{S}_n}) &\stackrel{{\rm{(a)}}}{=} \log \left| \{A \: : \: p(A|{Q_n=q_n,S_n=\mathcal{S}_n}) > 0 \} \right| \\
 &= \log \left| \bigcup_{\mathbf{g} \in  \mathcal{G}{({q_n,\mathcal{S}_n})}} \{ A\: : \:  \exists  \mathbf{d} \in \mathbb{F}_2^{T}, \mathbf{g} = \mathbf{d} A \} \right| \\
 &\geq \log \left| \{ A\: : \:  \exists  \mathbf{d} \in \mathbb{F}_2^{T}, \mathbf{g}^\prime = \mathbf{d} A \} \right| \\
 &\stackrel{{\rm{(b)}}}{\geq} \log \left| \{ L \subseteq \mathbb{F}_2^m \: : \:  \text{dim}(L) = T, \mathbf{g}^\prime \in L \} \right| \\
 &\stackrel{{\rm{(c)}}}{=} \log \prod\limits_{j=1}^{T-1} \left( \frac{2^m - 2^{j}}{2^T - 2^{j}} \right) \\
 &\stackrel{{\rm{(d)}}}{\geq} \log \left( \frac{2^m - 2}{2^T - 2} \right)^{T-1}  
= \Omega \left( mT - T^2 \right) ,
\end{align*}
where: {(i) the equality in} ${\rm{(a)}}$ follows from {Property~3} of the MIL; {(ii) the inequality in ${\rm{(b)}}$ follows by letting $L \subseteq \mathbb{F}_2^m $ be} a subspace of dimension $\text{dim}(L)$;
{(iii) the equality in}
${\rm{(c)}}$ follows by using Lemma~\ref{lem::subspaceCount} with $k=1$ (since $\mathbf{g}^\prime$ has only one row) and $t=T$; 
{(iv) the inequality in} ${\rm{(d)}}$ follows by noting that $\left(\dfrac{2^m - 2^j}{2^T - 2^j} \right) \geq \left(\dfrac{2^m - 2}{2^T - 2} \right) $ for $j \in [T-1]$.

\section{Proof of Theorem~\ref{theorem_main} - Equation~\eqref{eq::lb} and Lemma~\ref{lemma::SpecCase}}
\label{app::theorem_main}
\noindent\textbf{Theorem~\ref{theorem_main} - Equation~\eqref{eq::lb}.}
Given an index coding matrix $\mathbf{A}$, we denote by {${V}_{\mathbf{A}} \subseteq \mathbb{F}_{2}^T$} the subspace formed by the span of the rows of $\mathbf{A}$. 
It is {clear that} the dimension of {${V}_{\mathbf{A}}$} is at most $T$ (exactly $T$ {if $\mathbf{A}$ is full rank)} and that the $n$ distinct rows of $\mathbf{G}$ lie in {${V}_{\mathbf{A}}$.}
Let $\mathbf{a}_i \in \mathbb{F}_2^{m}, i \in [T_k],$ be the $i$-th row of $\mathbf{A}_k$.
Then, the problem of finding a lower bound on the value of $T_k$ can be formulated as follows: \textit{what is a minimum-size set of vectors $\Set{A}_{k} = \{ \mathbf{a}_{[T_k]}\}$ such that any row vector of $\mathbf{G}$ can be represented by a linear combination of at most $k$ vectors of {$\Set{A}_{k}$}?}

A lower bound on $T_k$ can be obtained as follows.
Given $\Set{A}_{k}$, there must exist a linear combination of at most $k$ vectors of {$\Set{A}_{k}$} that is equal to each of the $n$ distinct row vectors of $\mathbf{G}$. 
The number of \textit{distinct} non-zero linear combinations of 
up to $k$ vectors 
is at most equal to $\sum\limits_{j=1}^k {T_k \choose j}$. 
Thus, we have
\begin{align}
\label{eq:LBRep}
  \sum\limits_{i=1}^{k} {T_k \choose i} \geq n.
\end{align}

Combining this with the fact that $T_k \geq T$ gives precisely the bound in~\eqref{eq::lb}.

\noindent\textbf{Lemma~\ref{lemma::SpecCase}.}
We now derive the lower bound in Lemma~\ref{lemma::SpecCase}. 
We first consider the case where $n = 2^T-1$.
From~\eqref{eq:LBRep}, we obtain
 \begin{align}
  \sum\limits_{i=1}^{k} {T_k \choose i} \geq 2^{T} - 1. \label{lem_2}
 \end{align}
{Since in general $T_k \geq T$, to prove that $T_k \geq T+1$ for $k<T$, it is sufficient to show that we have a contradiction for $T_k=T$.
Indeed, by setting $T_k=T$, the bound in~\eqref{lem_2} becomes
\begin{align*}
\sum\limits_{i=1}^{k} {T \choose i} \geq 2^{T} - 1 = \sum\limits_{i=1}^{T} {T \choose i},
\end{align*}
which clearly is not possible since $k<T$.
Hence, $T_k \geq T+1$ for all $k<T$.}

For a general $n$ and $1 \leq k < \left\lceil T/2 \right\rceil$, we have
\begin{align*}
  k \left(\dfrac{T_k e}{k} \right)^k &\geq k {T_k \choose k} \geq \sum\limits_{i=1}^{k} {T_k \choose i} \geq n \label{lem_2} \\
   {\implies} T_k &\geq \dfrac{k^{\frac{k-1}{k}}}{e}  n^{1/k} = \Omega(k {n^{\frac{1}{k}}}).
 \end{align*}
Therefore, $T_k = \Omega(k 2^{\frac{T}{k}})$ when $n = \Theta(2^T)$.
 This lower bound, along with the upper bound in equation~\eqref{theorem_ub1} concludes the proof of Lemma~\ref{lemma::SpecCase}.

\bibliographystyle{IEEEtran}
\bibliography{Bib_v2}
%
%

\end{document}